\documentclass[letterpaper, 10 pt, conference]{ieeeconf}  
\usepackage{amsmath}
\usepackage{amsfonts}
\usepackage{graphicx}
\usepackage{url}

\IEEEoverridecommandlockouts  

\title{\LARGE \bf
Multi-Compartment Volume Conductor with Complete Electrode Model: Simulated Stereo-EEG Source Localization using Brainstorm-Zeffiro Plugin
}

\author{Fernando Galaz Prieto$^{1}$,  Takfarinas Medani$^{2}$, Chinmay Chinara$^{2}$,  Richard M.\ Leahy$^{2}$, and Sampsa Pursiainen$^{1}$
\thanks{This study was supported by the Research Council of Finland through the Center of Excellence in Inverse modeling and Imaging 2018--2025 (353089) and the  Flagship of Advanced Mathematics for Sensing, Imaging and modeling (FAME) (359185).}
\thanks{$^{1}$Fernando Galaz Prieto ({\tt\small fernando.galazprieto@tuni.fi}) and Sampsa Pursiainen ({\tt\small sampsa.pursiainen@tuni.fi}) are with Mathematics Research Center, Computing Sciences, Tampere University, Tampere, Finland.}
\thanks{$^{2}$Takfarinas Medani ({\tt\small medani@usc.edu}), Chinmay Chinara ({\tt\small chinara@usc.edu}) and Richard M.\ Leahy ({\tt\small leahy@usc.edu}) are with the Signal \& Image Processing Institute, University of Southern California, Los Angeles, CA USA.}
}

\begin{document}

\maketitle
\thispagestyle{empty}
\pagestyle{empty}

\begin{abstract}
This study introduces a novel integration of the Brainstorm (BST) software and the Zeffiro Interface (ZI) to enable whole-head, multi-compartment volume conductor modeling for electroencephalography (EEG) source imaging, with a particular focus on stereotactic EEG applications. We present the BST-2-ZI plugin, a MATLAB-based tool that facilitates seamless transfer of tissue segmentations and anatomical atlases from BST into ZI for finite element (FE) mesh generation as well as forward and inverse modeling. The generated FE meshes support variable spatial resolution and implement the complete electrode model (CEM), allowing for precise modeling of both invasive depth electrodes and non-invasive scalp electrodes. Using the ICBM152 template and synthetic source simulation, we demonstrate the end-to-end pipeline from MRI data to lead field (LF) computation and source localization in a stereotactic EEG (stereo-EEG) setting. Our numerical experiments highlight the capability of the pipeline to accurately model multi-compartment head geometry and conductivity  with a stereotactic CEM-based electrode configuration. Our preliminary source localization results show how a synthetic stereo-EEG probe corresponding to  a bidirectional deep brain stimulation (DBS) probe with four omnidirectional contacts can, in principle, be coupled with scalp electrodes to improve source localization in its vicinity.
\end{abstract}

\section{Introduction}
\label{sec:intro}

This numerical study aims to advance whole-head multi-compartment volume conductor modeling with variable spatial resolution for localizing human brain activity. Advanced numerical whole-brain modeling is an important topic when invasive depth-electroencephalography (depth-EEG) measurements are combined with non-invasive scalp-EEG or magnetoencephalography (MEG) data \cite{knosche2022eeg, hamalainen1993magnetoencephalography}. Namely, in such a system, both invasive EEG electrodes and their surroundings need to be modelled in conjunction with the subject's whole head, which calls for adapting the spatial resolution of the numerical discretization locally within the volume conductor. In particular, we aim to show how using an unstructured multi-compartment tetrahedral finite element (FE) mesh which incorporates complete electrode model (CEM) boundary conditions \cite{pursiainen2012complete}, one can successfully discretize the whole head with a stereotactic electrode setting and perform source localization in the near vicinity of the electrodes. 

EEG and MEG source localization represent an inverse problem, where the goal is to estimate the brain's primary current density from measurement data by inverting a linear forward model described by the lead field (LF) matrix. This inversion must be performed using data that typically includes noise. Solving the inverse problem requires modeling the subject’s head as a volume conductor. For accurate solutions across the entire brain, it is generally necessary to employ multi-compartment head models derived from magnetic resonance imaging (MRI)~\cite{vorwerk2014guideline, MEDANI_2023_119851, galaz2023multi}. These models capture individual anatomical features and local variations in tissue properties, particularly electrical conductivity, which is essential for precisely mapping neural sources in clinical applications such as epilepsy source localization~\cite{salayev2006spike, lahtinen2024standardized}. The electric fields recorded in EEG are especially sensitive to conductivity changes, meaning that both the forward mapping and the inverse solution are strongly influenced by the structure of the volume conductor.

Several numerical methods have been explored to find the LF matrix, each with distinct advantages based on how they model head tissues and electrical conductance. Boundary Element Method (BEM) \cite{gramfort2011forward, mosher1999eeg} is commonly employed, assuming homogeneous conductivities across different regions. While the computational cost of BEM grows along with the number of compartments, we focus on the FE method \cite{miinalainen2019realistic, he2020zeffiro, pursiainen2017advanced} which allows for creating a head model with an arbitrary number of compartments, spatially varying conductivity and level of detail, as well as to solve the source localization problem for the whole brain \cite{rezaei2021reconstructing, lahtinen2024standardized}. 

We perform our numerical experiments using Brainstorm-Zeffiro pipeline BST-2-ZI for creating a FE mesh-based volume conductor model using the anatomical data imported from  Brainstorm (BST) \cite{tadel2011brainstorm} to another open MATLAB-based toolbox Zeffiro Interface's (ZI's) \cite{he2020zeffiro} and its FE mesh generation routine \cite{galaz2023multi}. To demonstrate the capabilities of BST-2-ZI, we use BST's average brain template ICBM152 (International Consortium for Brain Mapping) \cite{ lancaster2007bias}. The workflow spans from importing the ICBM152 template from BST into ZI to generating the LF matrix and solving the source localization problem for a synthetic source using ZI. We demonstrate that our numerical framework enables us to accurately localize synthetic thalamic dipoles using a whole-head FE-based stereotactic EEG setting combining non-invasive scalp and invasive depth electrodes. Our implementation has been integrated as part of the MATLAB-based forward and inverse processing tool in ZI platform, supporting both FE meshing and simulated EEG source imaging.

\section{Methods and Materials}
\label{sec:MM}

We begin by briefly deriving the weak formulation of the EEG forward problem under CEM boundary conditions and how the LF matrix formulation can be obtained from it via FE discretization. CEM assumes that a portion of the external boundary \( \partial \Omega \) of the head domain \( \Omega \), modeled as a volume conductor, is covered by electrodes. Each electrode corresponds to a surface patch \( e_{\ell} \) for \( \ell = 1, 2, \ldots, L \), with contact area \( |e_{\ell}| \) and electrode potential \( U_{\ell} \). The governing equation for the electric potential distribution inside \( \Omega \) is derived from the principle of charge conservation, which states that the divergence of the total current density \( \vec{J} \) vanishes within \( \Omega \), i.e., 
\[
\nabla \cdot \vec{J} = \nabla \cdot\left(\vec{J}^p - \sigma \nabla u\right) = 0 \quad \text{in} \quad \Omega.
\]
Here, \( u: \Omega \to \mathbb{R} \) denotes the electric potential field and volume conduction, \( \sigma: \Omega \to \mathbb{R}^{3 \times 3} \) denotes the electrical conductivity tensor field, and \( \vec{J}^p \) denotes the primary current density. This leads to the Poisson-type equation:
\begin{equation}
\nabla \cdot(\sigma \nabla u) = \nabla \cdot \vec{J}^p \quad \text{in } \Omega.
\end{equation}

\subsection{CEM Boundary Conditions}
\label{sec:CEM}
The CEM boundary conditions are as follows:
\begin{alignat}{2}
    0  & = \sigma \frac{\partial u}{\partial n}(\vec{x}), & \quad \text{ in } & \quad \partial \Omega \backslash \cup_{\ell=1}^L e_{\ell} \label{eq:CEM_1} \\
    0 & = \int_{e_{\ell}} \sigma \frac{\partial u}{\partial n}(\vec{x}) d S, & \quad \text { for } & \quad  \ell=1,2, \ldots, L \label{eq:CEM_2} \\
    U_{\ell} & =u(x)+\tilde{Z}_{\ell} \sigma \frac{\partial u}{\partial n}(\vec{x}), & \quad \text { for } & \quad \ell=1,2, \ldots, \vec{x} \in e_{\ell} \label{eq:CEM_3}
\end{alignat}

The first condition (\ref{eq:CEM_1}) assumes no electrical currents flow through the parts of the scalp not covered by electrodes. The second condition (\ref{eq:CEM_2}) states that the total current passing through the $\ell$-th electrode equals zero. According to the third condition (\ref{eq:CEM_3}), the potential $U_{\ell}$ at the $\ell$-th electrode is given by the sum of the skin potential and the electrode–skin potential jump, expressed as $\tilde{Z}_{\ell} \sigma \frac{\partial u}{\partial n}(\vec{x})$, where $\tilde{Z}_{\ell}$ (in $\Omega \cdot \mathrm{m}^2$) denotes the pointwise effective contact impedance. For simplicity, the impedance is assumed to have the form $\tilde{Z}_{\ell} = Z_{\ell} |e_{\ell}|$, where $Z_{\ell}$ (in $\Omega$) is the average contact impedance  of the electrode and $|e_{\ell}|$ is the electrode's surface area. This leads to the integral expression $U_{\ell} = {|e_{\ell}|}^{-1} \int_{e_{\ell}} u \, dS$, indicating that $U_{\ell}$ is the sum of the mean potential over the electrode and the potential jump caused by the contact impedance. The reference potential is fixed by requiring $\sum_{\ell=1}^L U_{\ell} = 0$. Under these assumptions, the potential field $u \in \mathcal{S}$ is determined by solving the following weak formulation:
\begin{equation}
\label{eq:cem}
\begin{aligned}
    \int_{\Omega} \sigma \nabla u \cdot \nabla v \, dV = & -\int_{\Omega} (\nabla \cdot \vec{J}^p) v \, dV \\
    & + \sum_{\ell=1}^L \frac{1}{Z_{\ell} |e_{\ell}|^2} \left( \int_{e_{\ell}} u \, dS \right) \left( \int_{e_{\ell}} v \, dS \right) \\
    & - \sum_{\ell=1}^L \frac{1}{Z_{\ell} |e_{\ell}|} \int_{e_{\ell}} u v \, dS,
    \end{aligned}
\end{equation}
for all $v \in \mathcal{S} \subset H^1(\Omega)$, where $\mathcal{S}$ is a suitably defined subspace of the Sobolev space $H^1(\Omega)$, consisting of functions with square-integrable values and gradients. We discretize $\Omega$ using the finite element method with $\mathcal{S}$ spanned by continuous finite element basis functions.

\subsection{Derivation of the Lead Field Matrix}

The finite element discretization of the potential field $u$ and the primary current density $\vec{J}^p$ can be represented as
$u_{\mathcal{T}} = \sum_{i=1}^N z_i \psi_i$ and $\vec{J}^p_{\mathcal{T}} = \sum_{k=1}^M x_k \vec{w}_k$, where the coefficient vectors $\mathbf{z}$ $=$ $(z_1$, $z_2$, $\ldots$, $z_N)$ and $\mathbf{x}$ $=$  $(x_1$, $x_2$, $\ldots$, $x_M)$ are associated with scalar-valued basis functions $\psi_1, \ldots, \psi_N$ and vector-valued basis functions $\vec{w}_1, \ldots, \vec{w}_M$, respectively, both defined over the finite element mesh $\mathcal{T}$. These representations are substituted into the weak formulation, leading to the following structured linear system:
\[
\begin{pmatrix}
\mathbf{A} & -\mathbf{B} \\
-\mathbf{B}^T & \mathbf{C}
\end{pmatrix}
\begin{pmatrix}
\mathbf{z} \\
\mathbf{u}
\end{pmatrix}
=
\begin{pmatrix}
-\mathbf{Gx} \\
\mathbf{0}
\end{pmatrix},
\]
which arises from a Ritz-Galerkin finite element discretization of the weak form~(2) and must be solved for the unknown electrode voltage vector $\mathbf{u} = (U_1, U_2, \ldots, U_L)^\top$. The stiffness matrix $\mathbf{A} \in \mathbf{R}^{N \times N}$ is then assembled as
\[
a_{ij} = \int_{\Omega} \sigma \nabla \psi_i \cdot \nabla \psi_j \, dV + \sum_{\ell=1}^L \frac{1}{Z_{\ell}} \int_{e_{\ell}} \psi_i \psi_j \, dS, \quad \text{for } i,j \neq i'.
\]

The coupling and source matrices are explicitly defined by:
\[
\begin{aligned}
b_{i\ell} &= \frac{1}{Z_{\ell}} \int_{e_{\ell}} \psi_i \, dS, \\
c_{\ell\ell} &= \frac{1}{Z_{\ell}} \int_{e_{\ell}} dS, \quad c_{i\ell} = 0 \quad \text{for } i \neq \ell, \\
g_{ik} &= \int_{\Omega} (\nabla \cdot \vec{w}_k) \psi_i \, dV,
\end{aligned}
\]
yielding the matrices $\mathbf{B} \in \mathbf{R}^{N \times L}$, $\mathbf{C} \in \mathbf{R}^{L \times L}$, and $\mathbf{G} \in \mathbf{R}^{N \times M}$. The measurement vector $\mathbf{y} = \mathbf{R} \mathbf{u}$ consists of the electrode potentials referenced to their average. The reference projection matrix $\mathbf{R} \in \mathbf{R}^{L \times L}$ is constructed as
$r_{jj} = 1 - {1}/{L}$, $r_{ij} = -{1}/{L}$ for $i \neq j$, $j = 1, \ldots, L$. Solving the linear system for $\mathbf{u}$ leads to an explicit expression for the measured voltages ${\bf y} = {\bf L} {\bf x}$ in which ${\bf L}$ is the CEM LF matrix of the form
\[
\mathbf{L} = \mathbf{R} \left( \mathbf{B}^T \mathbf{A}^{-1} \mathbf{B} - \mathbf{C} \right)^{-1} \mathbf{B}^T \mathbf{A}^{-1} \mathbf{G}.
\]

\subsection{BST-2-ZI Pipeline}

\begin{figure}[h!]
    \centering
    \includegraphics[width=0.95\linewidth]{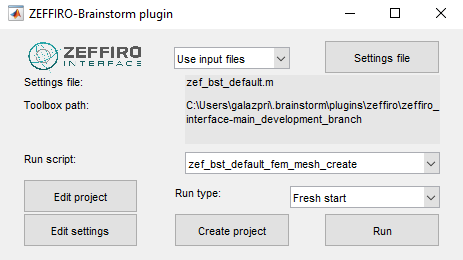}
    \caption{BST-2-ZI plugin dialog window as it appears when called from BST in a MATLAB environment. This interface allows the user to select the required input files for mesh generation—namely, the settings file and the run script—and to define the run mode (Fresh start, Import compartments, or Use project).}
    \label{fig:zef_plug}
\end{figure}

To transfer the data from BST to ZI, we use the BST-2-ZI plugin, which we have implemented as a part of ZI to allow generation of a tetrahedral FE mesh-based volumetric conductor model in ZI based on a tissue segmentation, a set of closed triangular surfaces obtained from the BST protocol\footnote{\url{https://neuroimage.usc.edu/brainstorm/Tutorials/CreateProtocol}}. For the user, BST-2-ZI appears as a simple dialog box (Fig.~\ref{fig:zef_plug}) which can be prompted when BST is running. BST-2-ZI is, for example, prompted by BST's FE mesh generation routine if ZI is chosen as the mesh generation platform. Preliminary user instructions can be found in \cite{brainstormFEM, zeffiroBST2ZI}. In BST, a typical full head segmentation includes the five main tissues: white matter (WM), gray matter (GM), CSF, skull, and scalp. In addition to these, BST-2-ZI allows importing BST protocol's MRI-based anatomical atlases as compartments in ZI. In BST's atlas, each voxel in the MRI is assigned to a specific tissue class. When importing an atlas to ZI, the voxel-based segmentation is converted into a set of surface meshes.

Integrating ZI with BST constitutes an important way to enhance model creation capabilities, since BST is extensively used by the neuroscience   community\footnote{\url{https://neuroimage.usc.edu/brainstorm/Community}}, and fully interfaced with several well-established segmentation software such as FastSurfer \cite{henschel2020fastsurfer}, FreeSurfer \cite{dale1999cortical}, SPM \cite{ashburner2005unified}, CAT12 \cite{gaser2016cat}, BrainSuite \cite{shattuck2002brainsuite}, CIVET \cite{ad2006civet}, and BrainVISA \cite{riviere2003brainvisa}. BST can either call some of these tools from the main interface for the anatomical segmentation or  import precomputed segmentation results. Note that only a T1-weighted (T1w) MRI data is required for the complete head segmentation, although T2-weighted (Tw2) data is optional but strongly recommended for better definition of the interface between skull and cerebrospinal fluid (CSF).  

After importing mesh segments as closed triangular surfaces, BST-2-ZI plugin (root path in ZI: \texttt{+utilities/+brainstorm2zef/}) allows the user to create an unstructured multi-compartment tetrahedral mesh through ZI's mesh creation routine (see Fig.~\ref{fig:zef_plug}). To produce the mesh, BST-2-ZI must be supplied a \textit{settings file} (default settings: \texttt{./+settings/zef\_bst\_default.m} w.r.t.\ root path) which lists user-defined parameters with their respective initial values; and \textit{run script} (\texttt{./+m/zef\_bst\_plugin\_start.m} w.r.t.\ root path) featuring a list of nested functions with explicitly declared inputs which define the instructions to be executed. A modal dialog box enables users to select these files as well as define at which point should the mesh generation shall commence based on the run type, which is one of the following: \textit{Fresh start} (BST's compartment data is used directly); \textit{Import compartments} (compartment data created by the previous run is imported); or \textit{Use project} (all data comes from a previously created ZI project).

Preliminary instructions for BST-2-ZI can be found in ZI's online repository\footnote{\url{\https://github.com/sampsapursiainen/zeffiro_interface/wiki/BST_2_ZI}}. Further information  can be found in BST's web pages\footnote{\url{https://neuroimage.usc.edu/brainstorm/Tutorials/FemMesh#Zeffiro}}.

\subsection{ICBM152 Template}

Brainstorm utilizes a version of the ICBM152 template \cite{lancaster2007bias} as its default anatomical model. This template can serve as a substitute when individual MRI scans are unavailable or unsuitable for processing or can be  employed as a common reference for group analyses. Brainstorm's default ICBM152 anatomy incorporates a tissue atlas  and parcellations derived from FreeSurfer's processing pipeline. These parcellations include the Desikan-Killiany atlas, which labels cortical regions based on gyral and sulcal anatomy, and Aseg atlas with subcortical structures segmented by FreeSurfer, providing a comprehensive anatomical framework for source localization and other analyses. 

\subsection{Numerical Experiments}

\begin{figure}[h!]
    \centering
    \begin{footnotesize}
    \begin{minipage}{4.2cm}
    \centering
    \includegraphics[height=3.5cm]{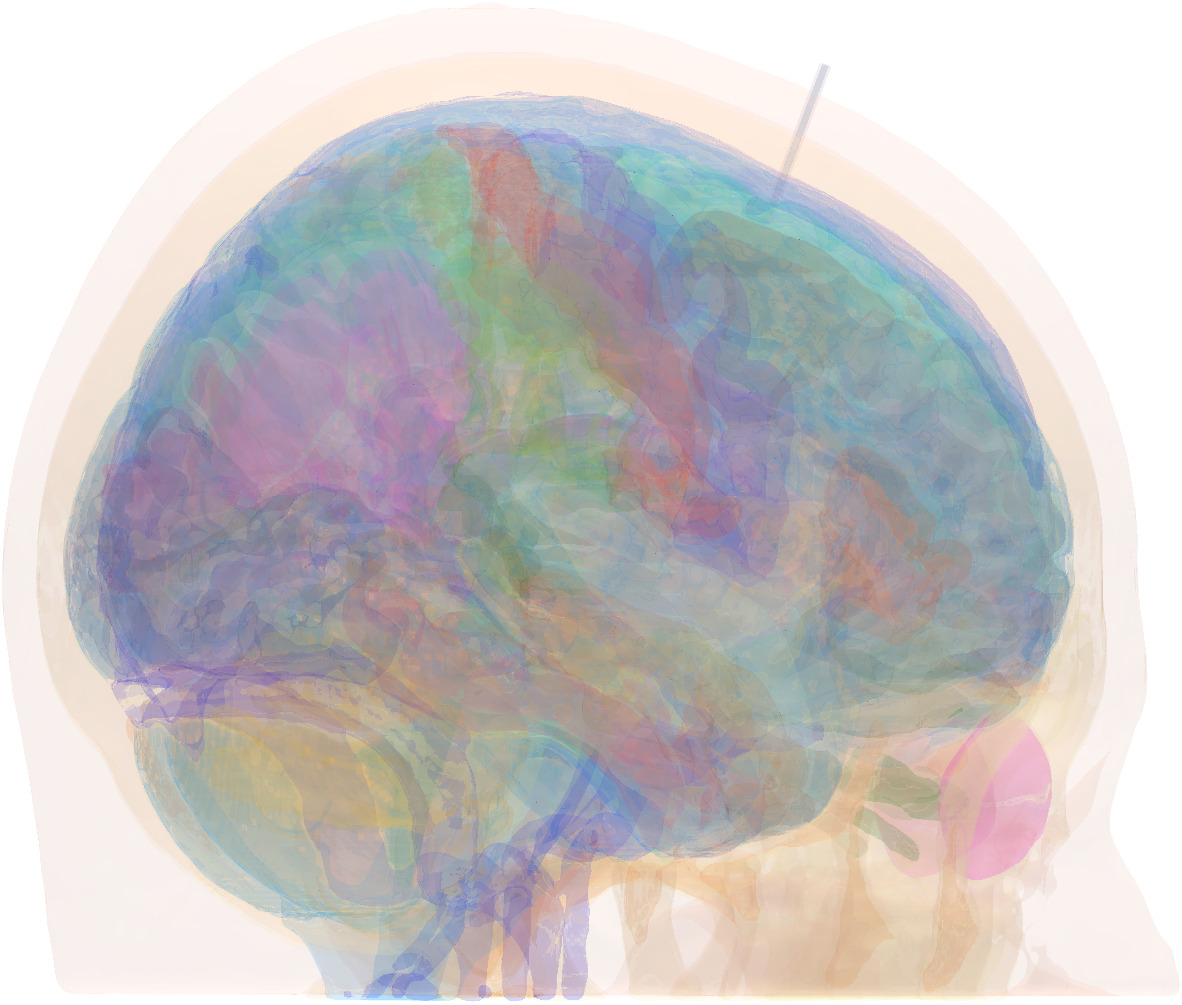} \\
    Transparent \\ surface \\ segmentation
    \end{minipage}
    \begin{minipage}{4.2cm}
    \centering
    \includegraphics[height=3.5cm]{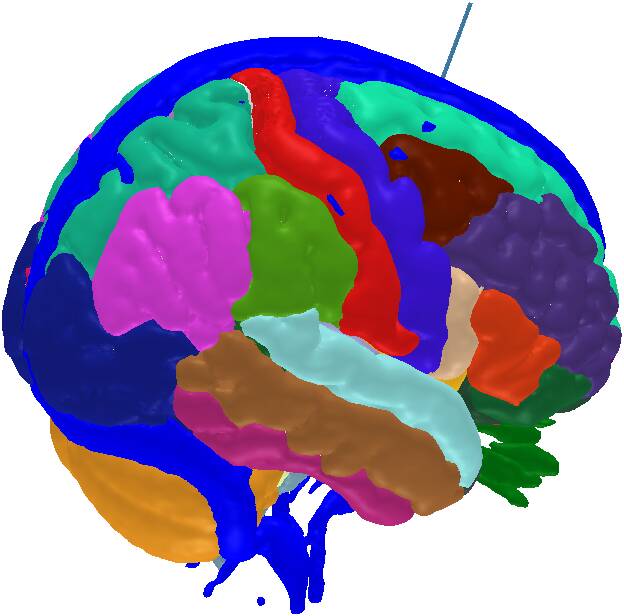} \\
    Surface mesh \\ compartments \\ inside skull
    \end{minipage}
    \end{footnotesize}
    \caption{A transparent surface segmentation of the whole head (left) with the stereo-EEG probe and the  compartments inside the CSF layer (right) as of after importing them from BST to ZI. Altogether, the number of compartments (scouts) imported from tissue atlas, Desikan-Killiany atlas and Aseg atlas, was 116. The electrical conductivities (S/m) of the different tissue types represented by those were chosen to be isotropic with the following values: white matter 0.14, gray matter 0.33, CSF 1.79, scalp 0.43, eyes 1.5, compact bone 0.0064, spongy bone 0.028, blood 0.7, muscle 0.33, probe 1E-15, and probe encapsulation 0.33. The white and gray matter compartments were composed of various sub-compartments of the Desikan-Killiany and Aseg atlas.}
    \label{fig:imported_segmentation}
\end{figure}

\begin{figure}[h!]
    \centering
    \begin{footnotesize}
    \begin{minipage}{4.2cm}
    \centering
    \includegraphics[height=2.6cm]{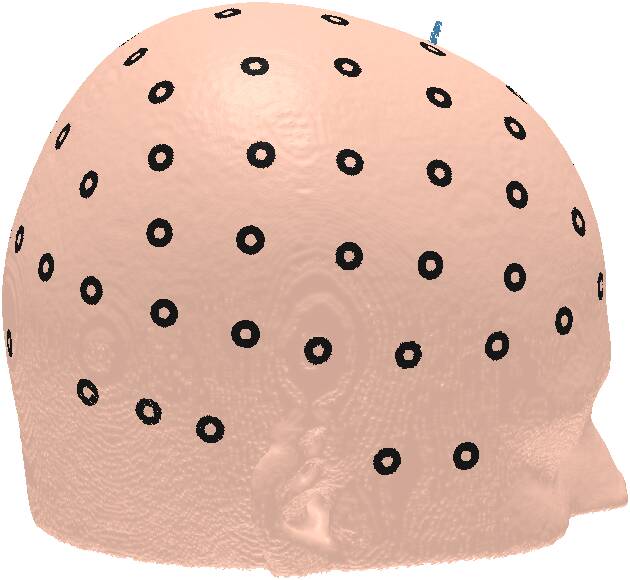} \\
    FE mesh with \\ scalp electrodes
    \end{minipage}
    \begin{minipage}{4.2cm}
    \centering
    \includegraphics[height=2.5cm]{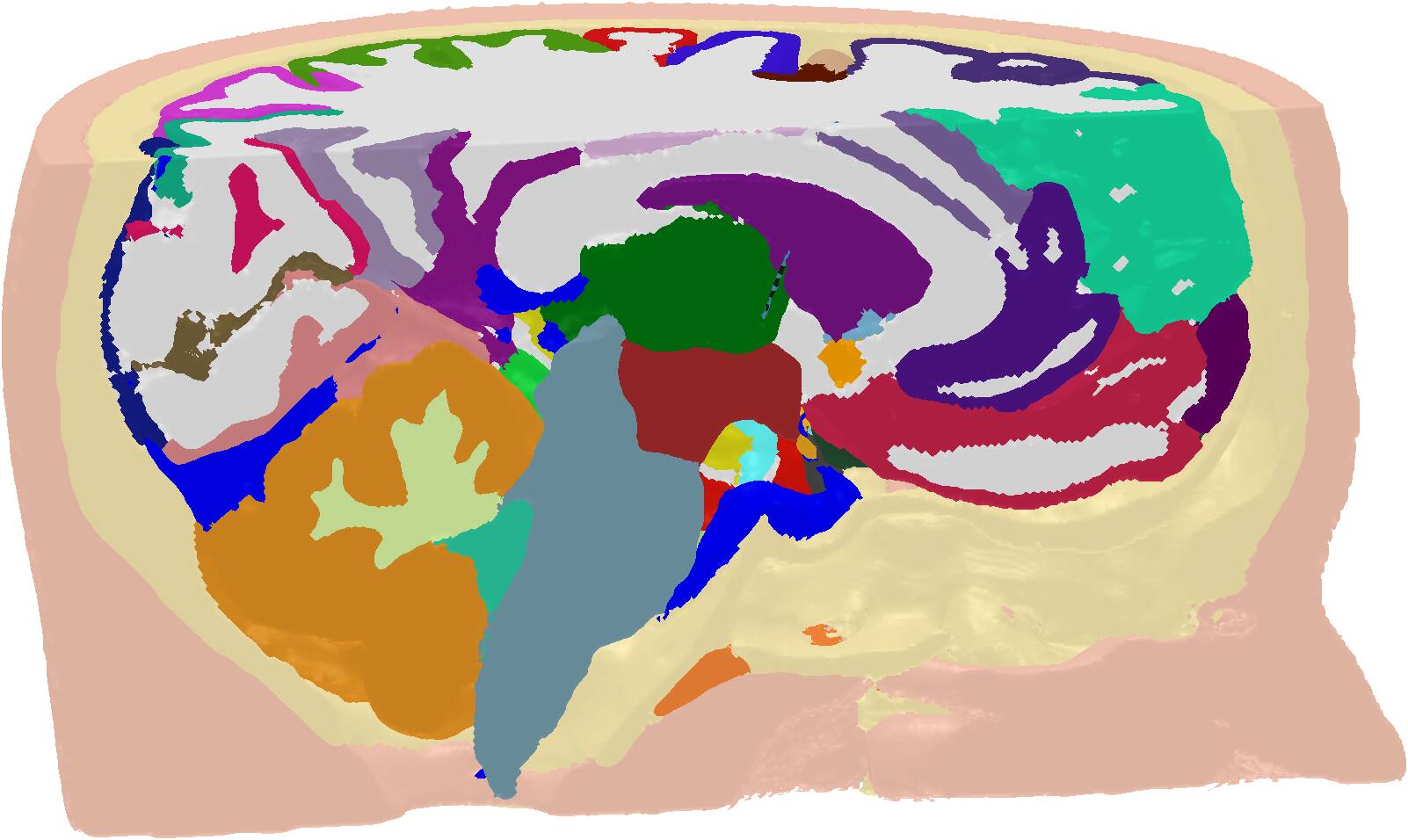} \\
    FE mesh without CSF 
    \end{minipage} 
    \end{footnotesize}
    \caption{A tetrahedral FE mesh with overall 0.6 mm resolution was generated as described in \cite{galaz2023multi} based on the 116 imported compartments together with 72 scalp electrodes, placed according to 10-20 system and a cylindrical stereo-EEG probe model with its 4 omnidirectional contacts (approximating those of Medtronic 3389, see, e.g.,  \cite{anderson2018optimized}) reaching the ANT region.}
    \label{fig:fe_mesh}
\end{figure}

\begin{figure}
    \centering
    \begin{minipage}{7cm}
    \centering
    \includegraphics[height=3.7cm]{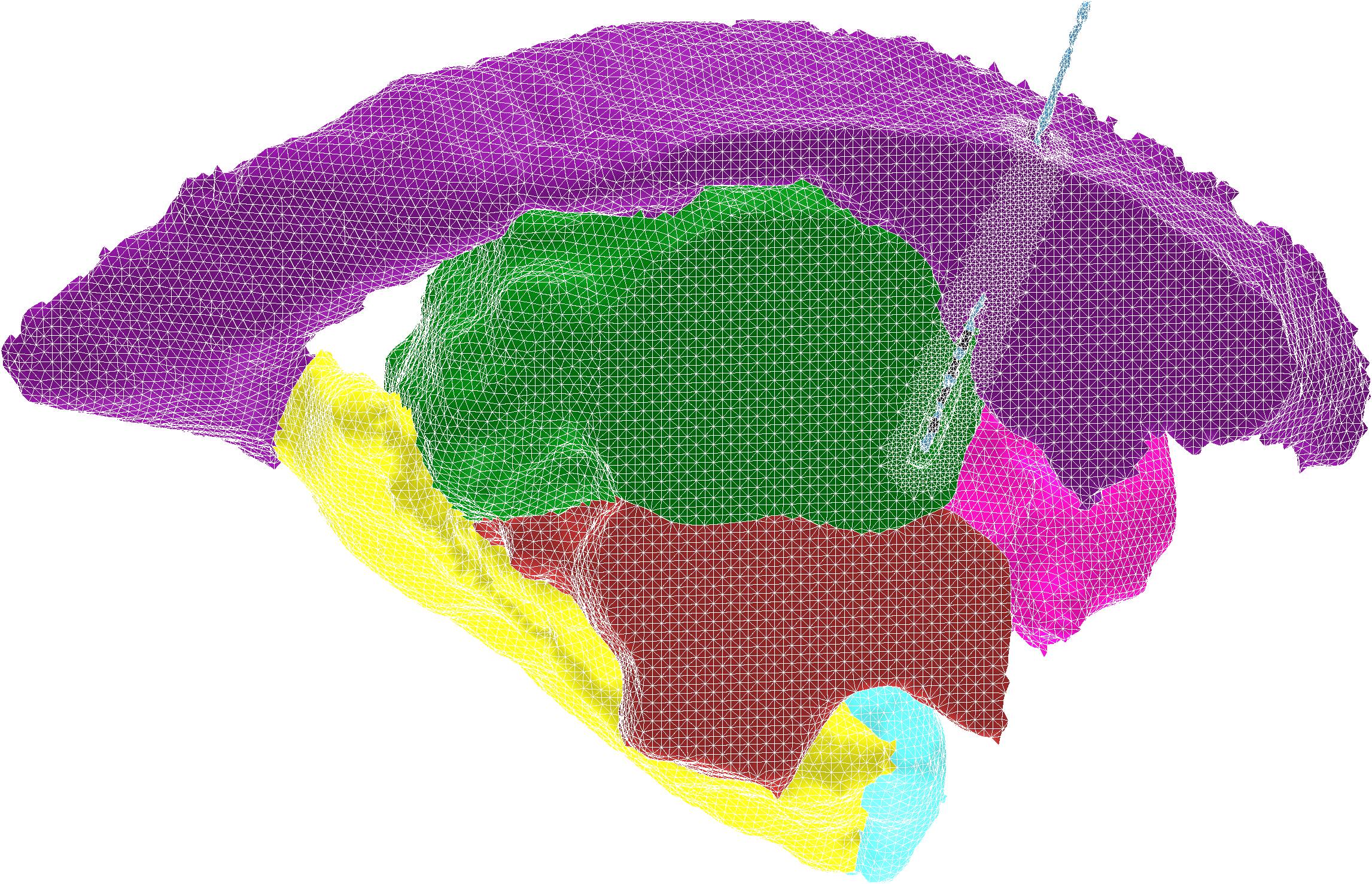} 
    \end{minipage}
    \caption{Deep structures with stereo-EEG probe; the FE mesh was refined in the vicinity of the probe to result in accuracy of approximately 0.3 mm with respect to the surface model. From Top to bottom: Ventricle Lateral (Dark purple), Putamen (Magenta), Hippocampus (Yellow), Amygdala (Cyan), Ventral Diencephalon (Red), Thalamus (Green).}
    \label{fig:thalamus_probe}
\end{figure}

\begin{table}[htbp]
    \caption{FE mesh specifications for the whole-head multi-compartment stereo-EEG model.}
    \centering
    \begin{tabular}{@{}|l|l|@{}}
    \hline
    \textbf{Parameter} & \textbf{Value / Description} \\
    \hline
    Mesh resolution (base) & 0.6 mm \\
    Mesh resolution (refined near probe) & 0.3 mm \\
    Number of compartments & 116  \\
    Total nodes & 18.7 million \\
    Total tetrahedra & 91.8 million \\
    Scalp electrodes & 72 (10–20 system) \\
    Stereo-EEG probe contacts & 4 (Medtronic 3389-like) \\
    Electrode model & Complete Electrode Model  \\
    \hline
    \end{tabular}
    \label{tab:fem_parameters}
\end{table}

The numerical setup for this study was established by first importing anatomical data from the ICBM152 template, specifically using the Desikan-Killiany and Aseg atlases, into ZI via Brainstorm (BST). This process yielded 116 segmented compartments, each represented by a closed surface mesh in Right-Anterior-Superior (RAS) orientation, as generated by the BST-to-ZI pipeline (Fig.\ \ref{fig:imported_segmentation}). Following this, 72 ring electrodes were positioned on the scalp in accordance with the 10-20 system. Each electrode had an outer diameter of 10.0 mm, an inner diameter of 4.0 mm, and a contact impedance of 2 k$\Omega$. Subsequently, a synthetic stereo-EEG probe was inserted approximately at the location of the anterior nuclei of the thalamus (ANT) using ZI’s strip tool—a graphical interface for implanting linear probes or electrode strips within the brain. The implanted cylindrical probe had a diameter of 1.27 mm and included four omnidirectional contacts, each with a length of 1.5 mm, an inter-contact spacing of 0.5 mm, and a contact impedance of 2 k$\Omega$. The geometry corresponds to the Medtronic 3389 device, commonly used in Deep Brain Stimulation (DBS) applications \cite{anderson2018optimized}, and suitable for stereo-EEG recordings in bidirectional modes \cite{anso2022concurrent}. To mimic the encapsulation typically observed in DBS procedures, the probe was enclosed within a concentric cylindrical layer of 0.5 mm thickness.

The FE mesh (Table \ref{tab:fem_parameters}) was created (Fig.\ \ref{fig:fe_mesh}) using ZI's mesh generator \cite{galaz2023multi} considering all 116 compartments. The electrical conductivities (S/m) of the different tissue types (Figure \ref{fig:imported_segmentation})  were chosen to be isotropic. The white and gray matter compartments were composed of various sub-compartments of the Desikan-Killiany and Aseg atlas.  In addition, to prevent void space inside the final volume conductor model, an envelope compartment with conductivity 0.33 S/m  was defined as a container of all other compartments and was associated with those FE mesh elements, if they otherwise were not labelled in the determination of the sub-compartments. 

The base resolution of the FE mesh was chosen to be 0.6 mm everywhere in the domain. The stereo-EEG probe and its encapsulation were refined uniformly to establish a sufficient mesh resolution of about 0.3 mm for modeling the electrodes on its surface (Figure \ref{fig:thalamus_probe}). The final mesh was composed of 18.7 M nodes and 91.8 M tetrahedra. 

To perform a simple source localization test using the whole-head electrode setting, two EEG LF matrices  were generated and tested: one for scalp-EEG and the other for stereo-EEG, corresponding to the combined configuration of scalp electrodes and stereo-EEG. The source space of both LF matrices was composed of altogether 100.000 source positions evenly distributed in the active compartments of the brain, including both cortical and subcortical areas, each one containing a triplet of three dipolar sources with Cartesian orientations. 

Synthetic data was generated using a single dipolar source placed in the region of the ventral posterior nucleus (VPN) in the left thalamus within 16 mm distance from the probe. The following two alternative orientations were examined: (i) close to parallel and (ii) close to perpendicular with respect to the probe. Gaussian zero-mean white noise with a 30 dB signal-to-noise ratio was used. As the source reconstruction techniques ZI's versions of sLORETA, Dipole Scan, and scalar Unit-Noise-Gain Beamformer (sUNGB) \cite{lahtinen2025mathematical} were applied.

\section{Results}

\begin{figure}[h!]
\centering
\begin{footnotesize}
       \begin{minipage}{7.0cm} 
\centering
        \includegraphics[height=2.8cm]{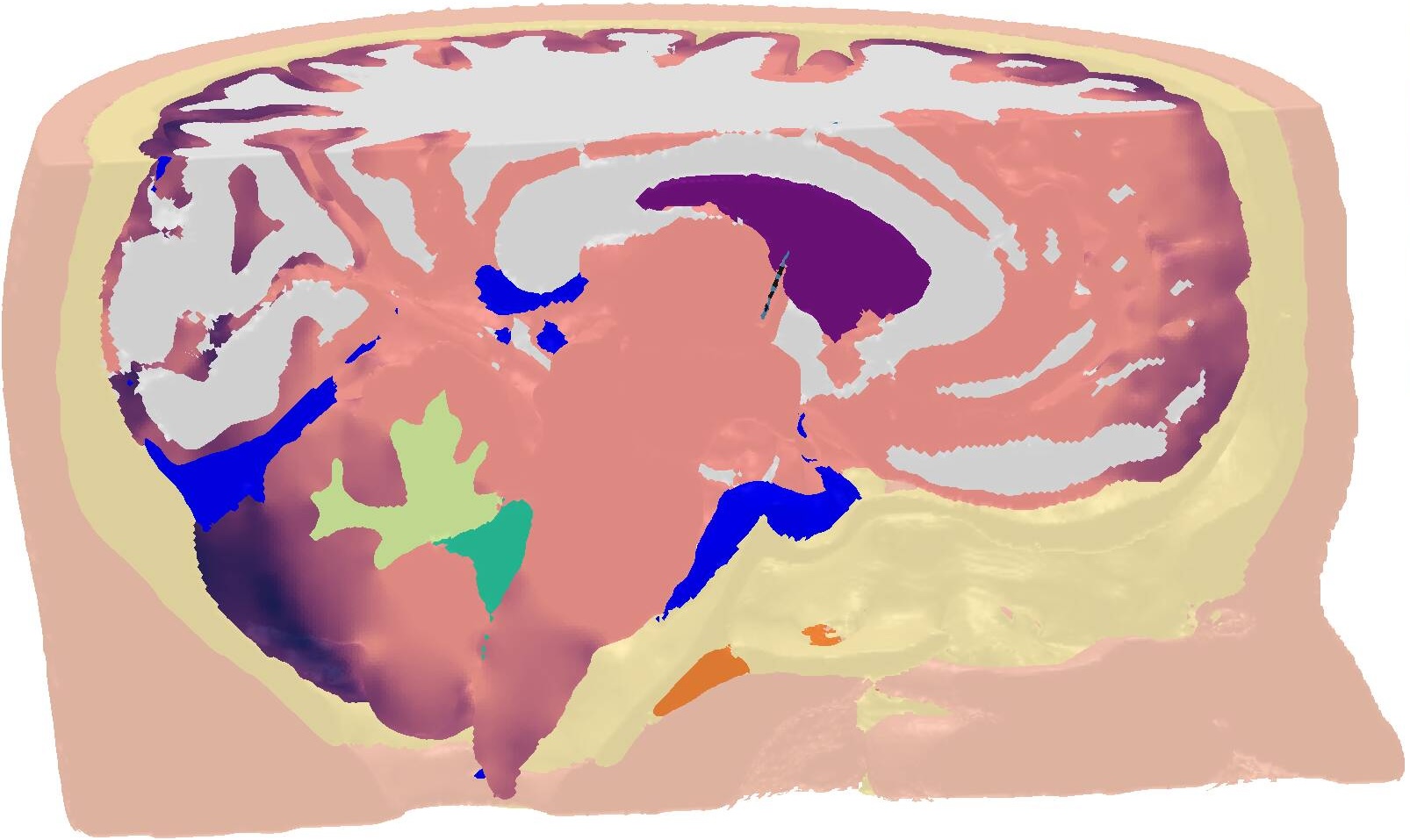} \hskip0.1cm \includegraphics[height=2.7cm]{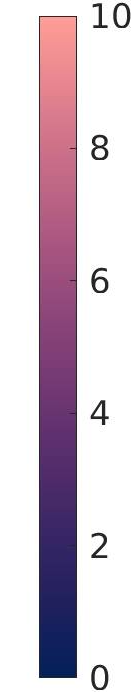}\\
\vskip0.1cm
LF for stereo-EEG \\
\mbox{}
           \end{minipage} 
       \begin{minipage}{7.0cm} 
\centering
        \includegraphics[height=2.7cm]{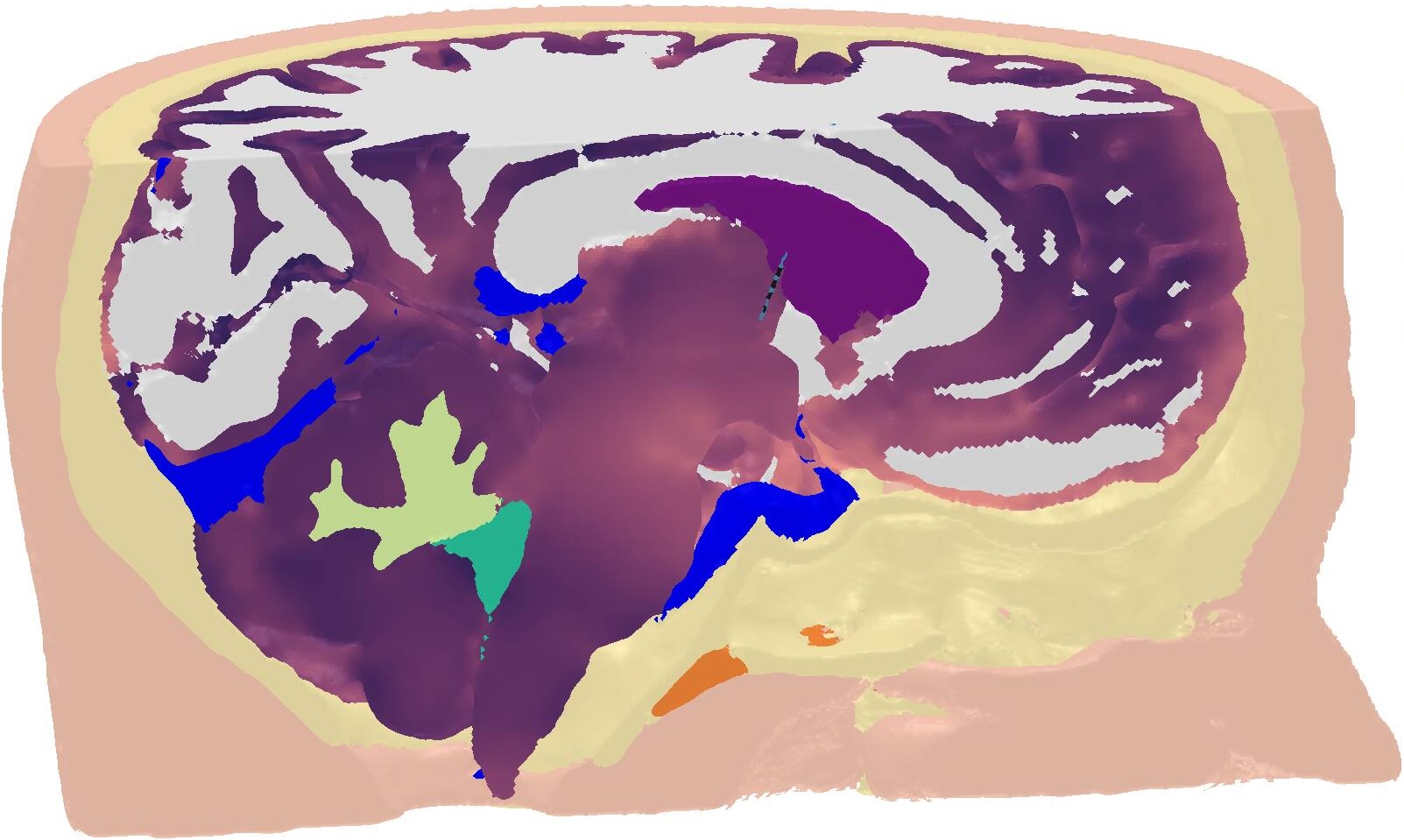} 
\hskip0.1cm \includegraphics[height=2.8cm]{color_bar.png}\\
\vskip0.1cm
LF for scalp-EEG \\
\mbox{}
           \end{minipage} 
           \end{footnotesize}
    \caption{The relative stereo LF amplitude for stereo-EEG and scalp-EEG as a volumetric distribution in decibels with respect to minimum value 0 dB and with 10 dB as a top threshold value. In the case of stereo-EEG the field can be observed to be concentrated into the vicinity of the  probe, while for scalp-EEG, the amplitudes are observed to be more evenly distributed with a growth towards the scalp electrodes.}
    \label{fig:lead_field}
\end{figure}

In the numerical experiments, two LF matrices were first generated using the whole-head multi-compartment FE mesh with two different CEM-based electrode settings; one of the matrices corresponds to scalp-EEG only and the other one to the combined stereo-EEG configuration.  Fig.~\ref{fig:lead_field} displays the relative spatial amplitude of the LF matrix for both scalp-EEG and combined stereo-EEG settings between 0~dB (visualization threshold) and 40~dB (maximum amplitude). In the stereo-EEG case, the LF  distribution is concentrated around the stereo-EEG probe, demonstrating the probe's ability to confine sensitivity to its immediate surroundings. In contrast, the scalp-EEG LF is more evenly spread and exhibits a larger overall footprint, with sensitivity increasing toward the scalp.

\begin{figure}[h!]
    \centering
    \begin{footnotesize}
        \begin{minipage}{1.8cm}
    \centering
    \includegraphics[trim={0 0 2cm 0},clip,height=1.5cm]{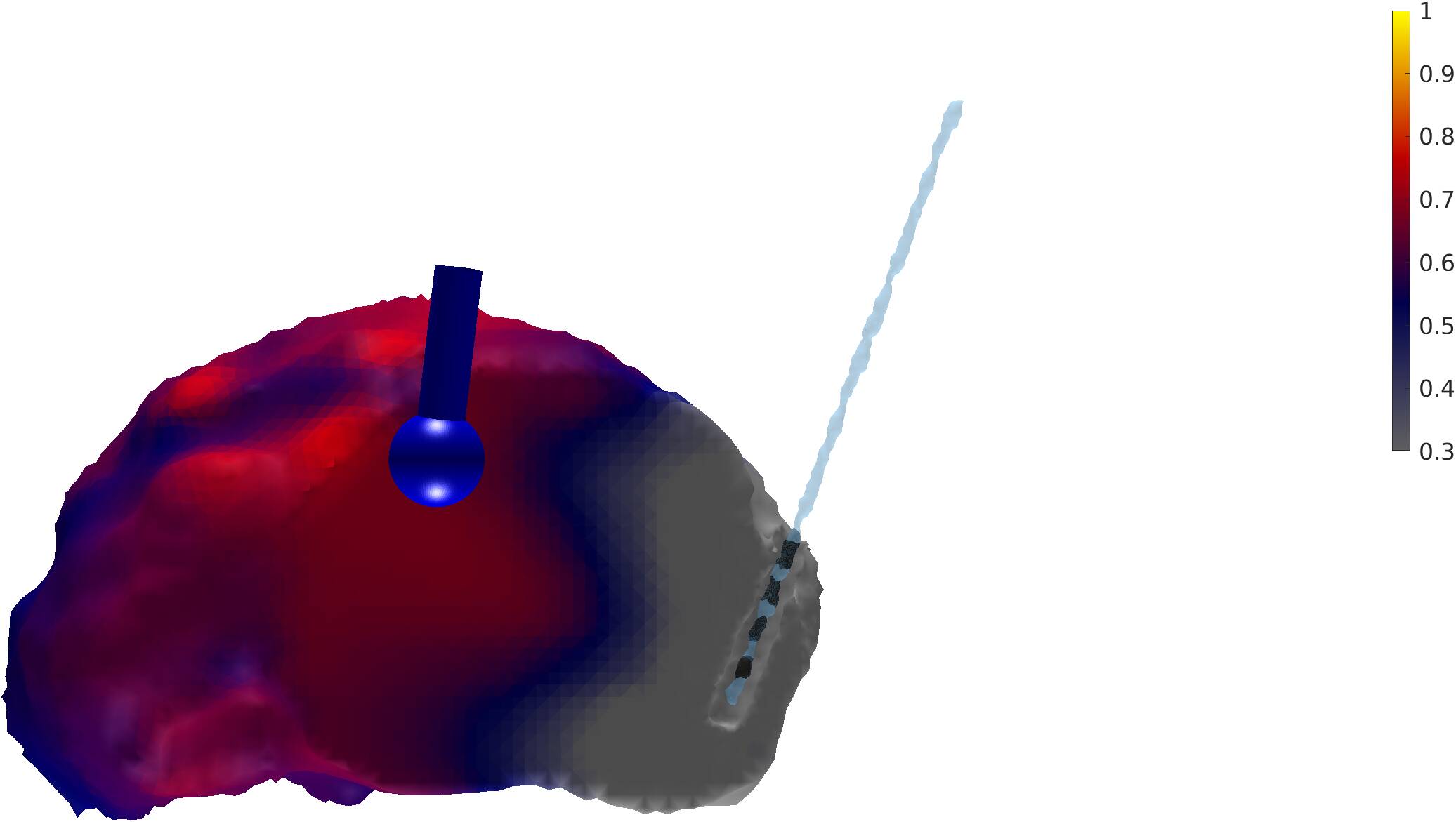} \\
    sLORETA, \\ Orientation (i)
    \end{minipage}
        \begin{minipage}{1.8cm}
    \centering
    \includegraphics[trim={0 0 2cm 0},clip,height=1.5cm]{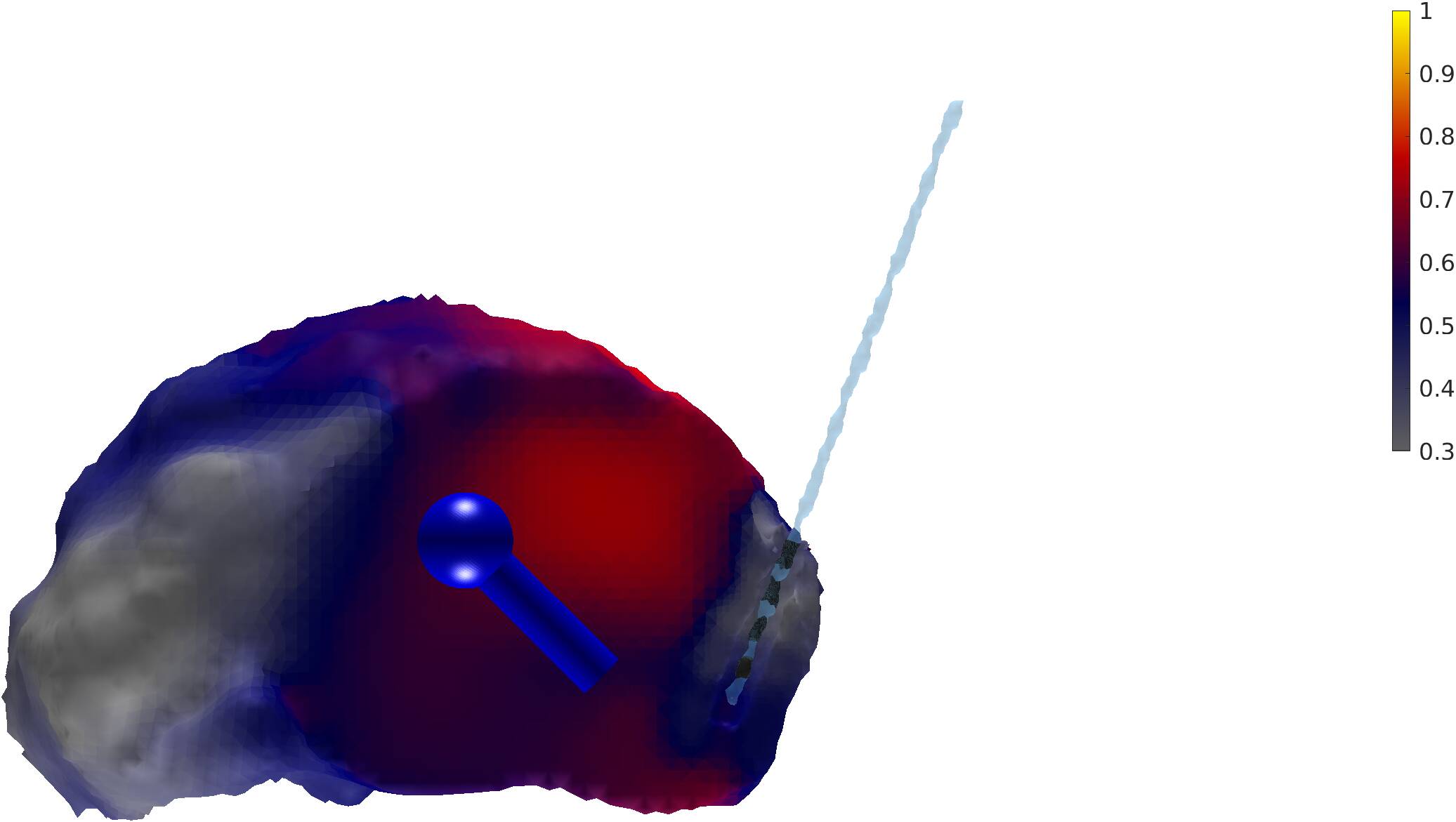}  \\ 
     sLORETA, \\ Orientation (ii)
    \end{minipage}
        \begin{minipage}{1.8cm}
    \centering
    \includegraphics[trim={0 0 2cm 0},clip,height=1.5cm]{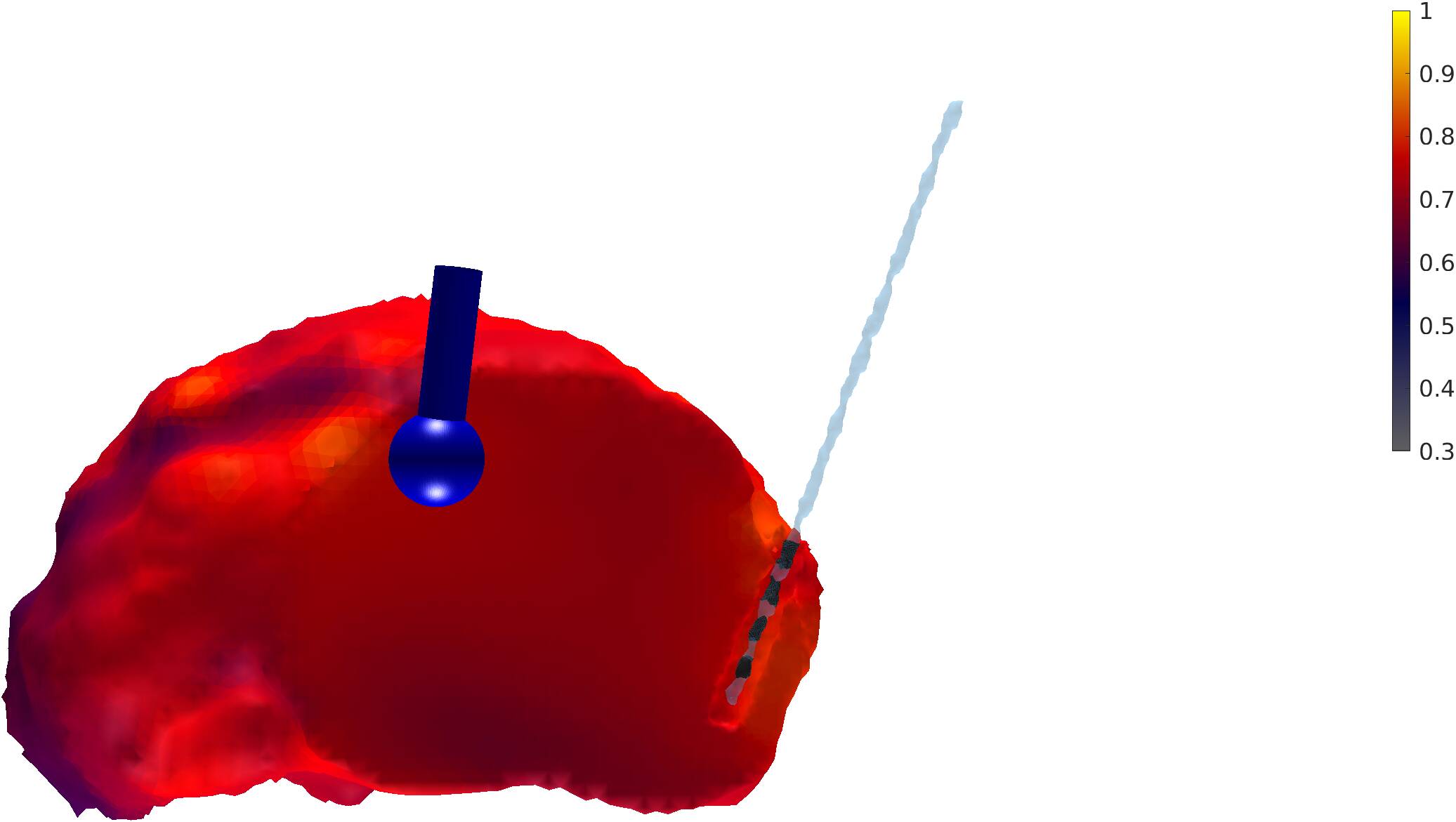}  \\ 
     sLORETA, \\ Orientation (i)
    \end{minipage} 
     \begin{minipage}{1.8cm}
    \centering
    \includegraphics[trim={0 0 2cm 0},clip,height=1.5cm]{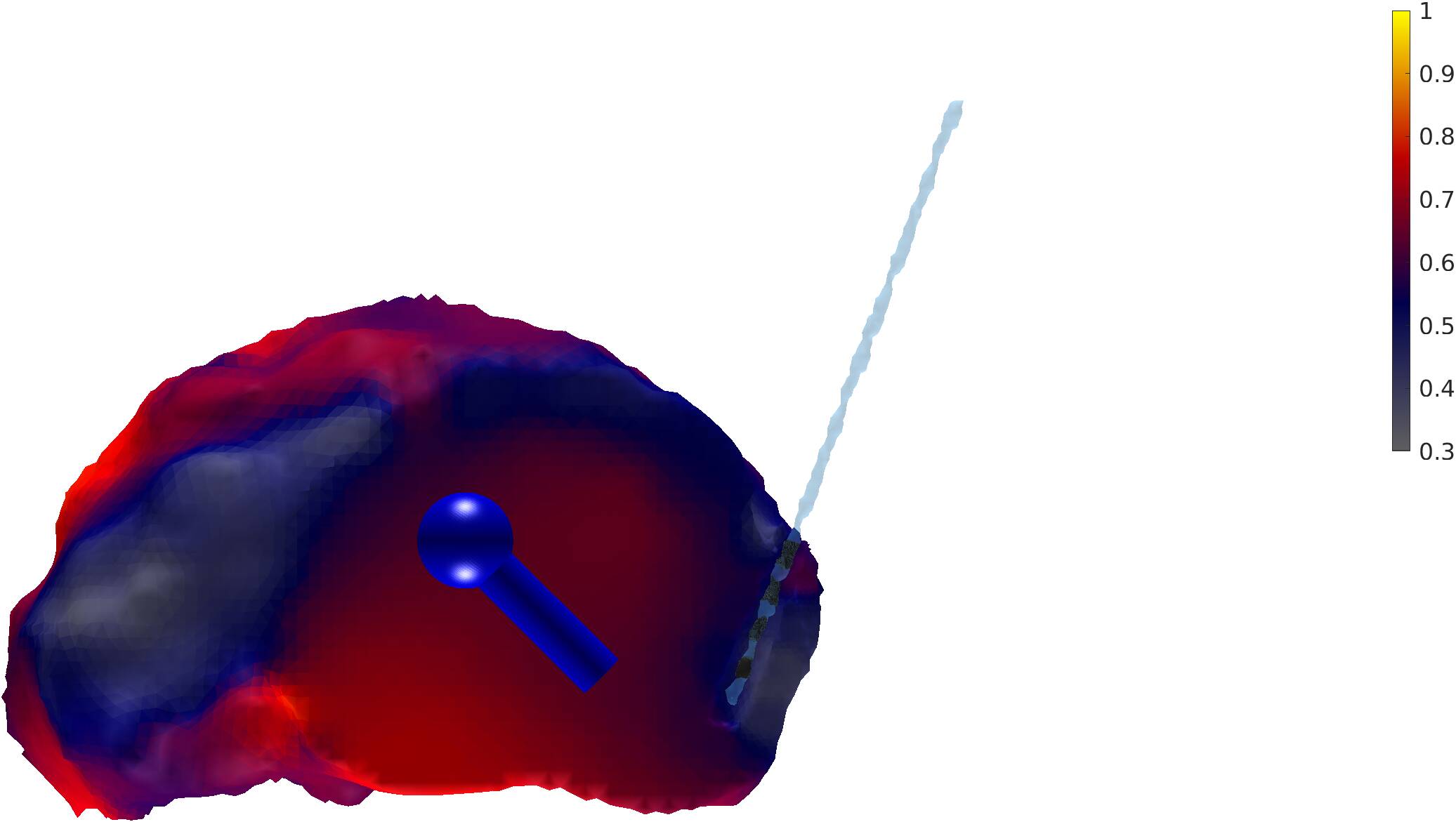}  \\ 
     sLORETA, \\ Orientation (ii)
    \end{minipage} 
       \begin{minipage}{0.3cm}
    \centering
 \includegraphics[trim={25cm 6cm 0 0},clip,height=2.2cm]{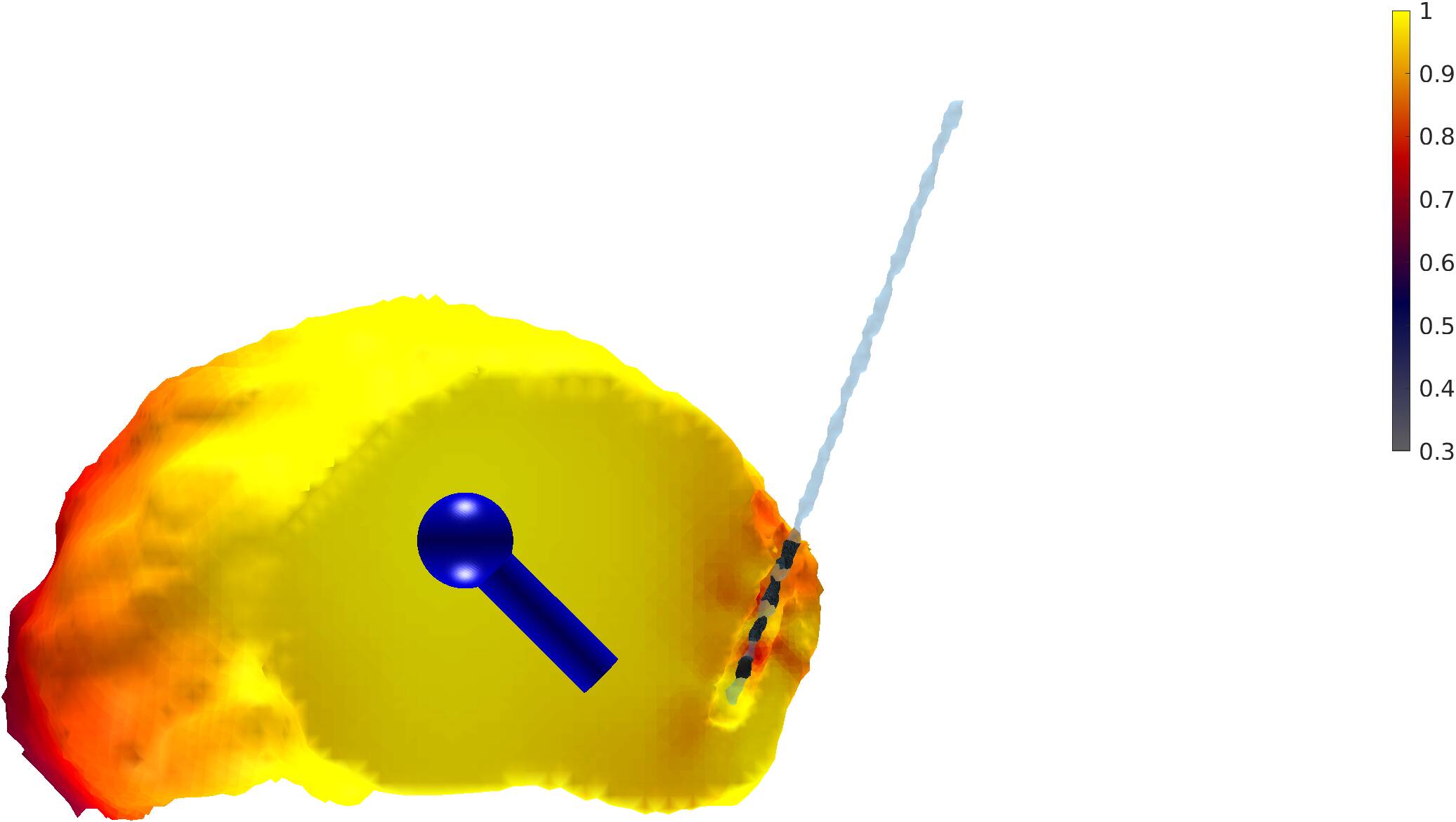}  
           \end{minipage}  
            \begin{minipage}{1.8cm}
    \centering
    \includegraphics[trim={0 0 2cm 0},clip,height=1.5cm]{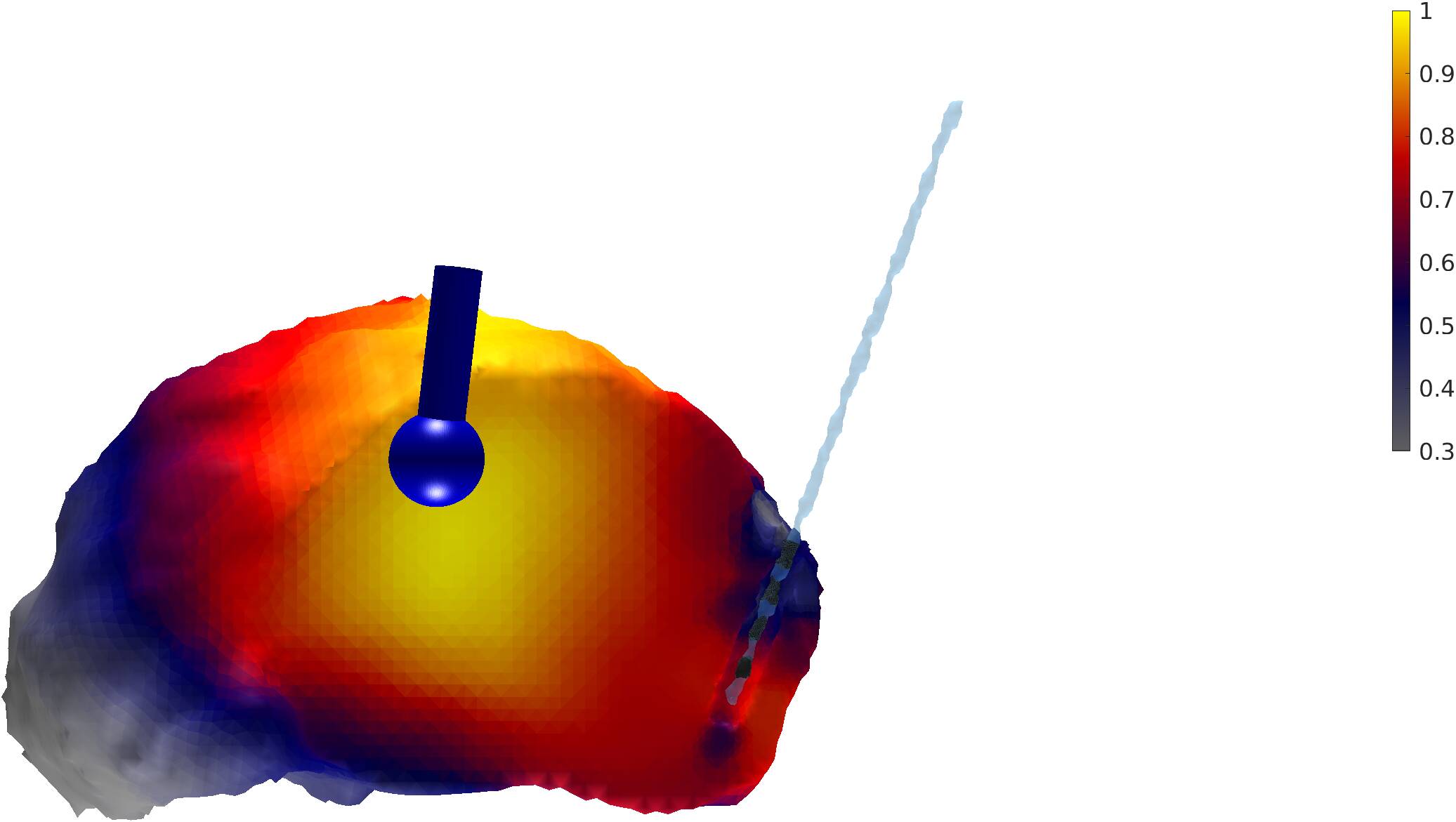} \\
    Dipole Scan, \\ Orientation (i)
    \end{minipage}
        \begin{minipage}{1.8cm}
    \centering
    \includegraphics[trim={0 0 2cm 0},clip,height=1.5cm]{thalamus_dipole_scan_2.jpg}  \\ 
     Dipole Scan, \\ Orientation (ii)
    \end{minipage}
        \begin{minipage}{1.8cm}
    \centering
    \includegraphics[trim={0 0 2cm 0},clip,height=1.5cm]{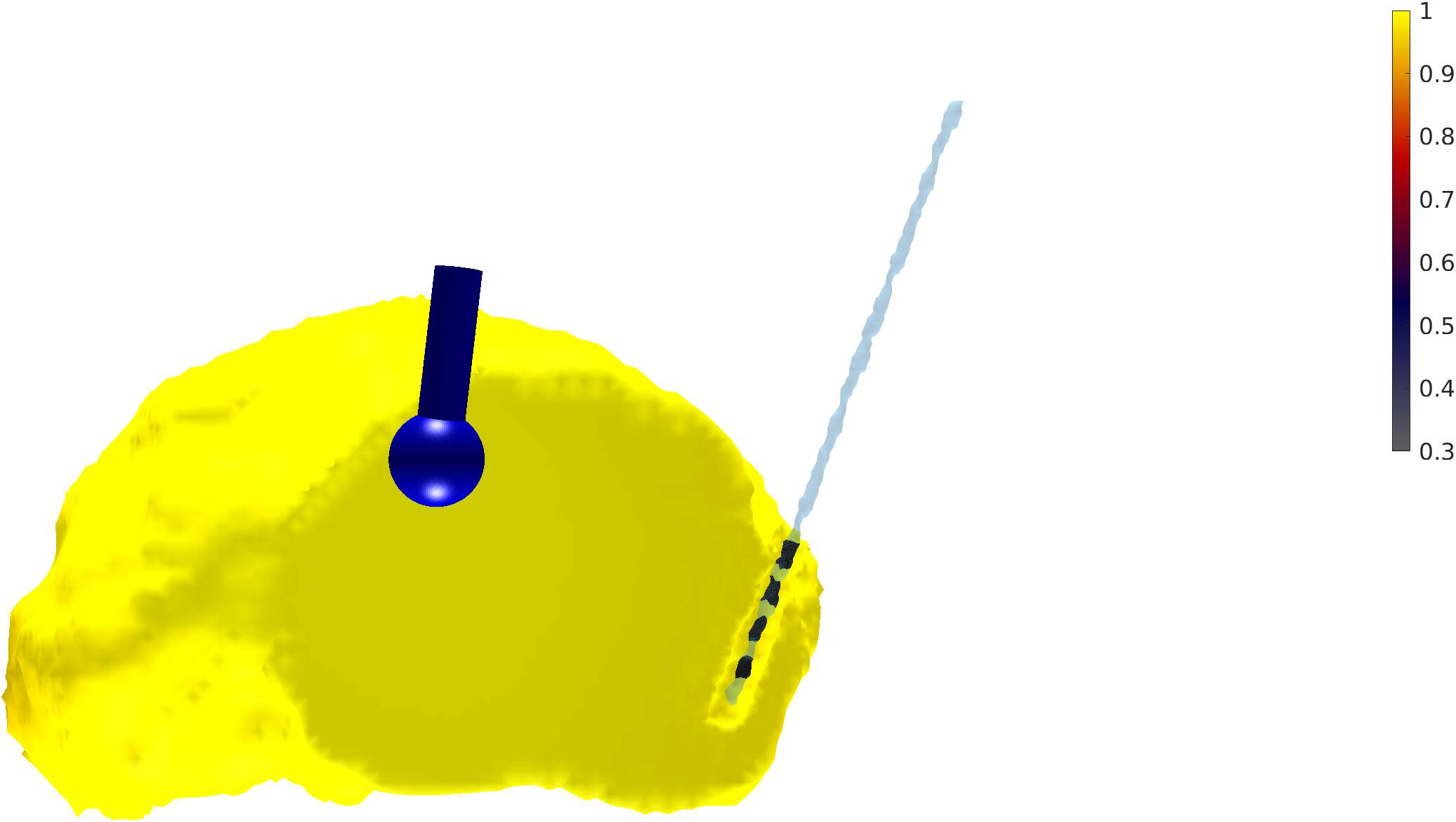}  \\ 
     Dipole Scan, \\ Orientation (i)
    \end{minipage} 
                \begin{minipage}{1.8cm}
    \centering
    \includegraphics[trim={0 0 2cm 0},clip,height=1.5cm]{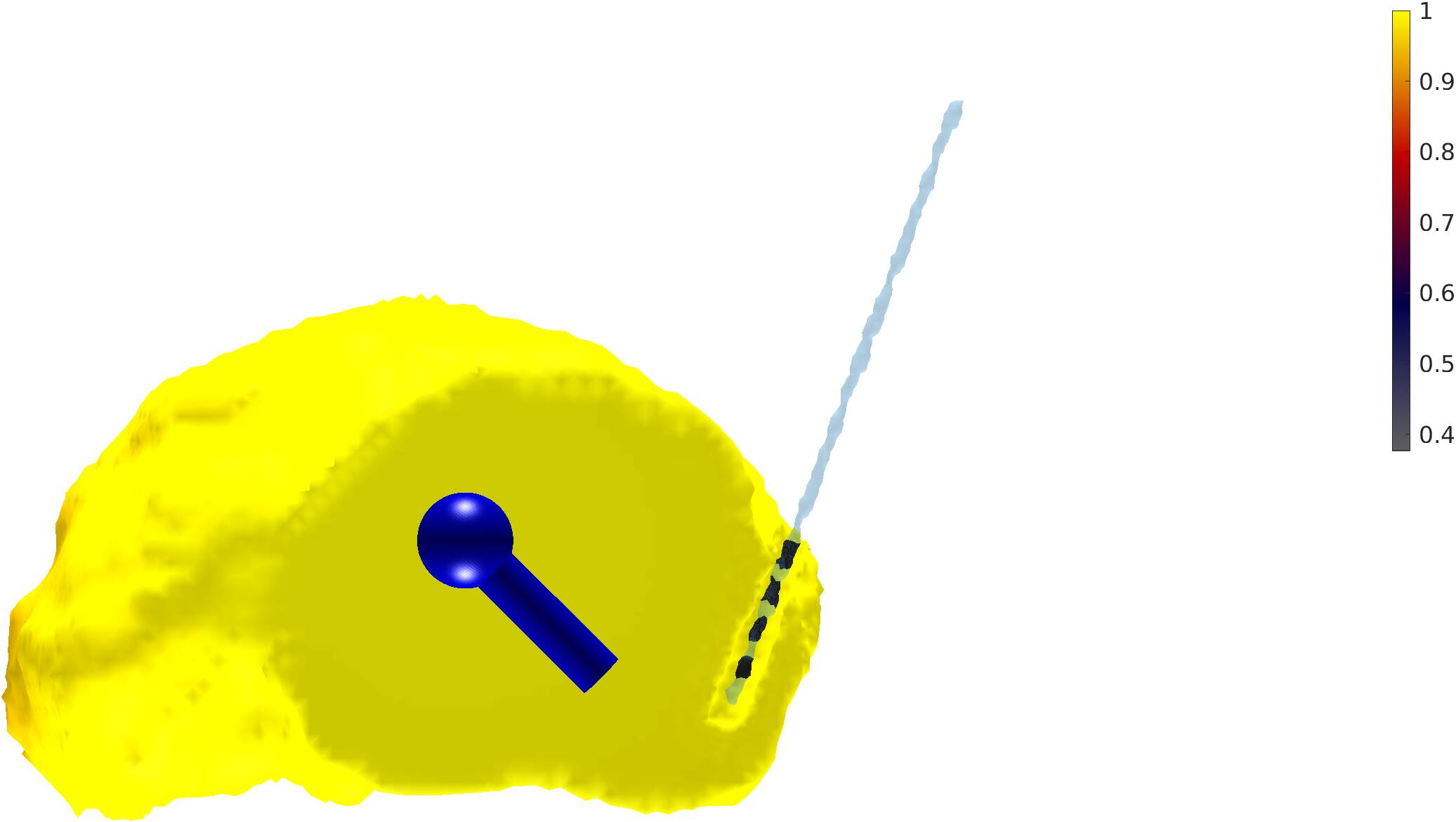}  \\ 
     Dipole Scan, \\ Orientation (ii)
    \end{minipage} 
    \begin{minipage}{0.3cm}
    \centering
   \includegraphics[trim={25cm 6cm 0 0},clip,height=2.2cm]{thalamus_dipole_scan_2.jpg}            
   \end{minipage}  
    \begin{minipage}{1.8cm}
    \centering
    \includegraphics[trim={0 0 2cm 0},clip,height=1.5cm]{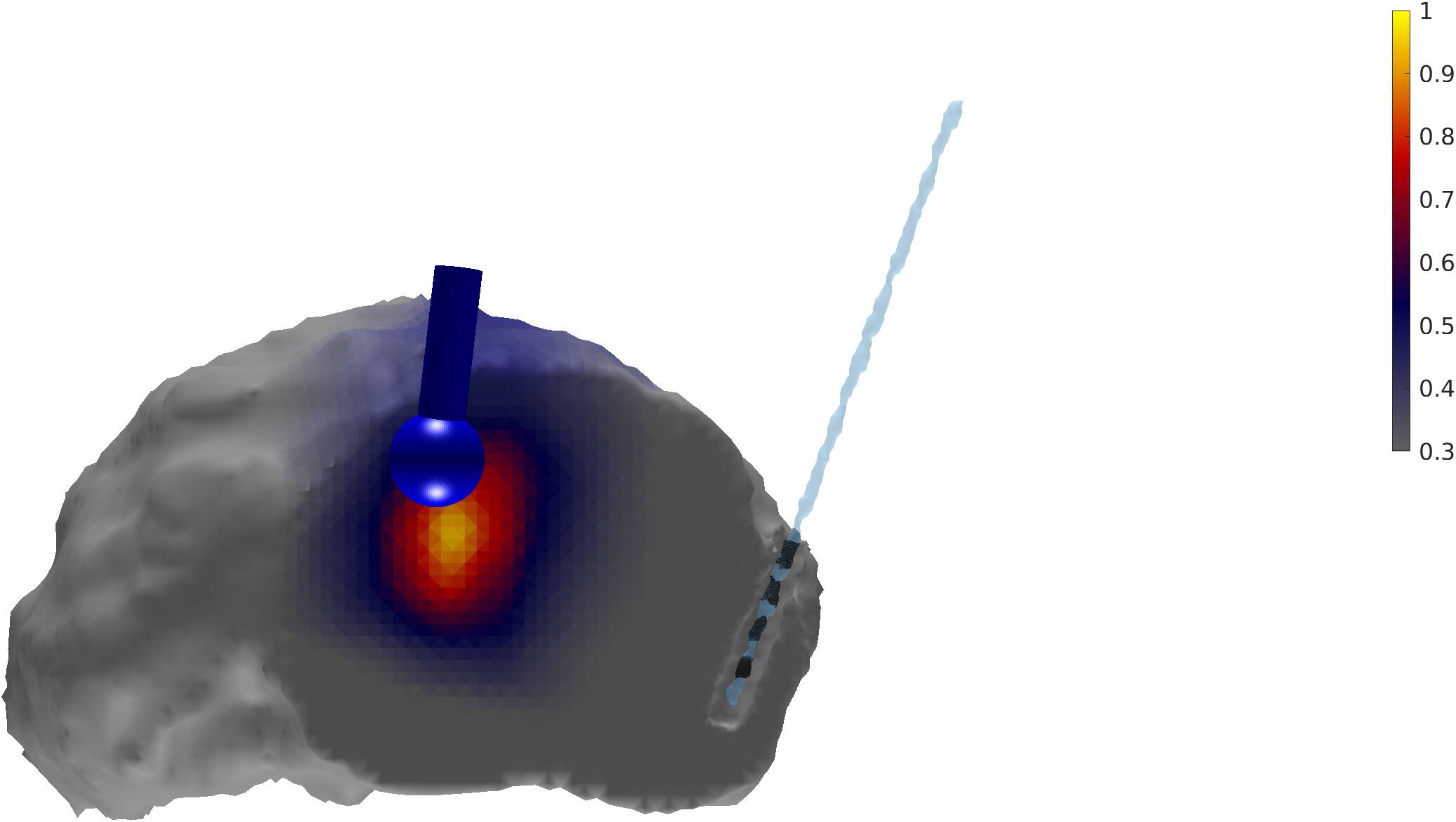} \\ 
    sUNGB, \\ Orientation (i)
    \end{minipage}
        \begin{minipage}{1.8cm}
    \centering
    \includegraphics[trim={0 0 2cm 0},clip,height=1.5cm]{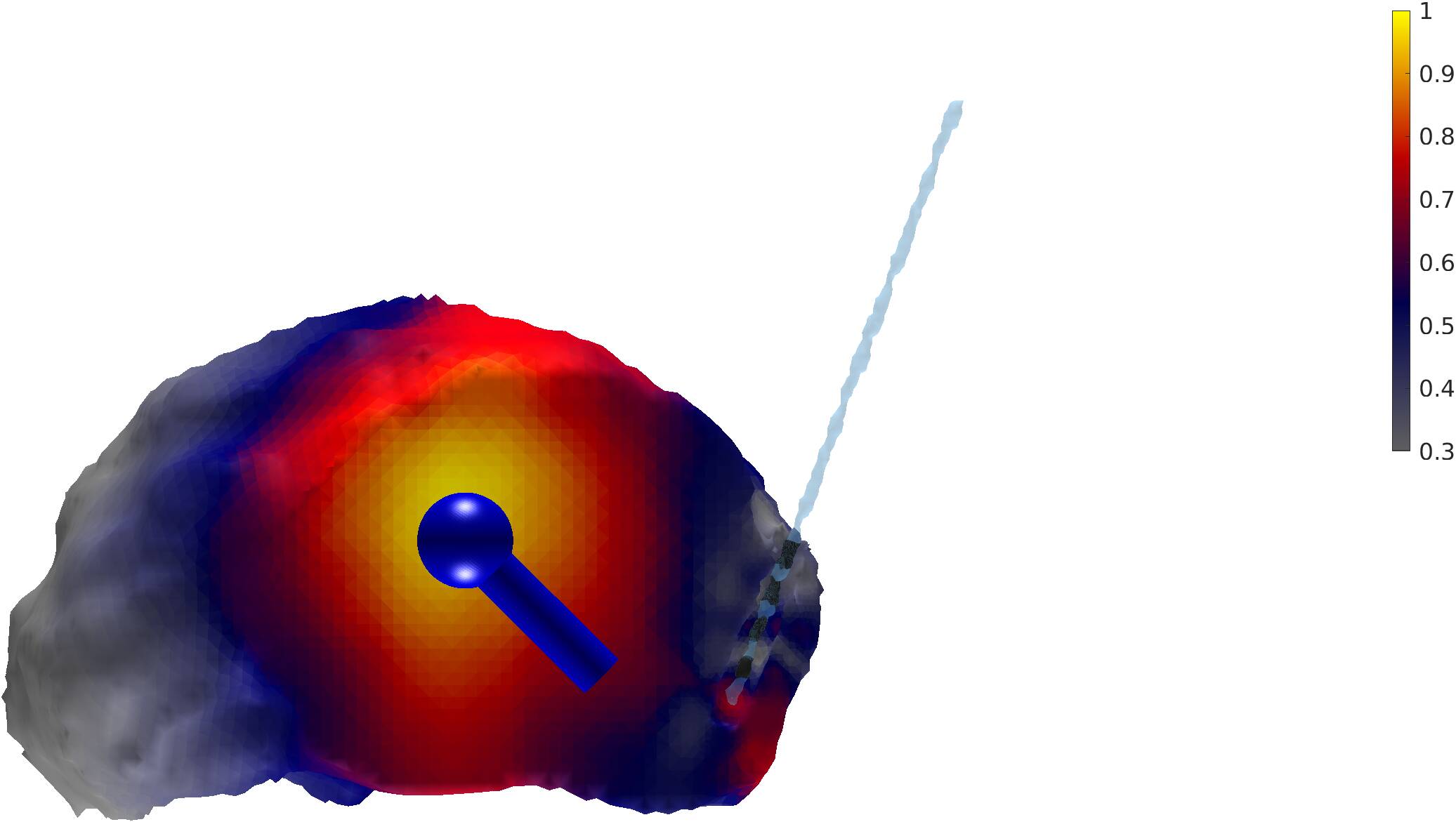} \\
    sUNGB, \\ Orientation (ii)
    \end{minipage}
        \begin{minipage}{1.8cm}
    \centering
    \includegraphics[trim={0 0 2cm 0},clip,height=1.5cm]{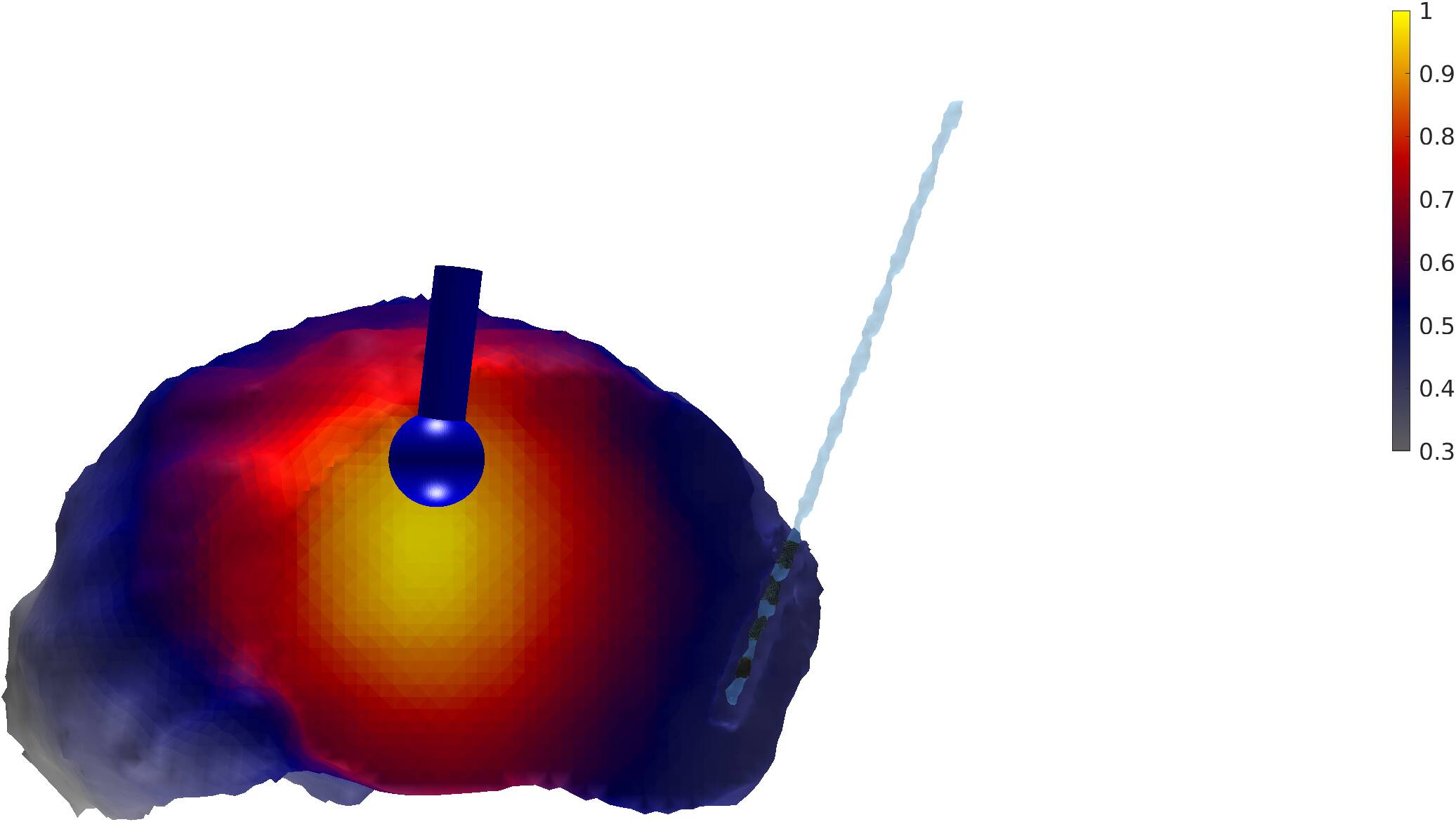}  \\ 
     sUNGB, \\ Orientation (i)
    \end{minipage}
        \begin{minipage}{1.8cm}
    \centering
    \includegraphics[trim={0 0 2cm 0},clip,height=1.5cm]{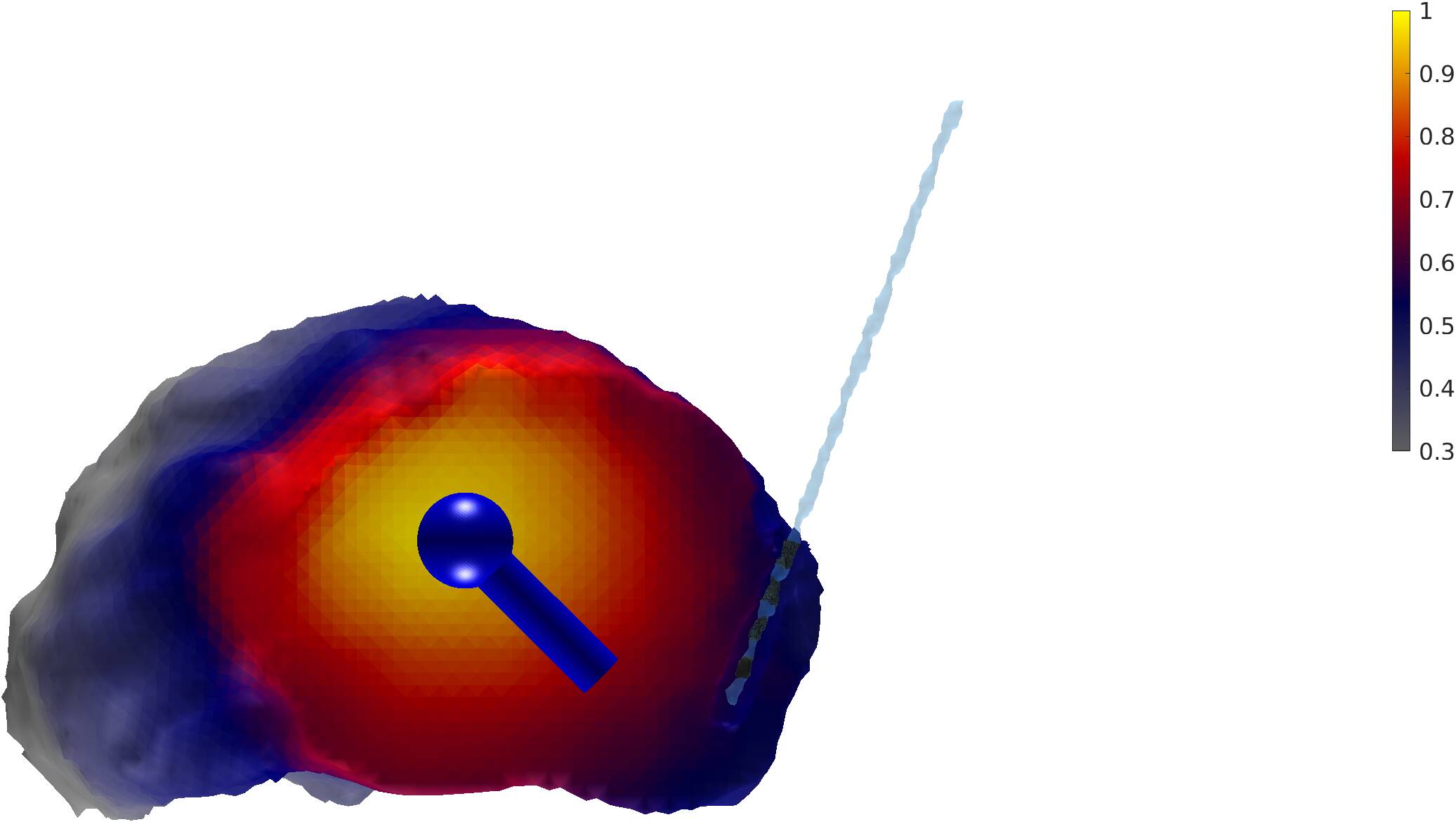} \\ 
    sUNGB, \\ Orientation (ii)
    \end{minipage}
       \begin{minipage}{0.3cm}
    \centering
   \includegraphics[trim={25cm 6cm 0 0},clip,height=2.2cm]{thalamus_dipole_scan_2.jpg}            
   \end{minipage}  \\ \vskip0.2cm
   \begin{minipage}{3.6cm}
   \centering
   Stereo-EEG
   \end{minipage}
     \begin{minipage}{3.6cm}
   \centering
   Scalp-EEG
   \end{minipage}
   \begin{minipage}{0.3cm}
   \centering 
   \mbox{}
   \end{minipage}
    \end{footnotesize}
    \caption{The source localization results obtained for a synthetic dipole with 30 dB SNR placed in the VPN region of the left thalamus with two alternative orientations: (i) close-to-parallel and (ii) close-to-perpendicular w.r.t.\ stereo-EEG probe. The sLORETA, Dipole Scan, and sUNGB were used as the source localization techniques and two whole-head LF matrices, one for scalp-EEG and the other one for stereo-EEG, as alternative forward mappings. The stereo-EEG LF included the 4 omnidirectional contacts of the probe in the configuration. Illustrated is the relative amplitude of the source reconstruction obtained in thalamus. The results demonstrate that, overall the distributions are more focused in the case of the stereo-EEG LF matrix. In the presence of the probe, the close-to-parallel orientation (i) yields a more focused estimate than the close-to-perpendicular one (ii). This tendency is not observed if the probe is absent. Dipolar source placement and orientation in VPN thalamus corresponds to the blue pin.}
    \label{fig:source_localization}
\end{figure}

The two LF matrices were applied to localize two dipolar sources in the left thalamus.  Fig.~\ref{fig:source_localization} shows the source localization results using sLORETA, Dipole Scan, and sUNGB for two orthogonal orientations of a synthetic thalamic dipole placed in the VPN region of the left thalamus. For all three reconstruction techniques, the results demonstrate that the inclusion of the stereo-EEG setting substantially improves the localization focus in the thalamic region. Specifically, when the dipole orientation is nearly parallel to the stereo-EEG probe, the spatial accuracy and concentration of the reconstructed source are consistently higher than for the perpendicular orientation. This phenomenon is evident for all three methods, with the effect being most pronounced in the sUNGB reconstructions. 

In contrast, in the absence of the probe electrodes (scalp-EEG setting), the localization results are generally more diffuse and show less dependence on dipole orientation. This suggests that the inclusion of invasive electrodes provides not only enhanced spatial precision but also sensitivity to the source orientation relative to the probe. While sUNGB yields the most focal results overall, it also appears to be the least affected by the presence or absence of the stereo-EEG probe in the perpendicular case.

\section{Discussion}

This study demonstrated a preliminary step toward anatomically accurate whole-head FE modeling for stereotactic EEG source localization. By defining electrodes via the CEM boundary conditions, we could create a whole-head system, in which both scalp electrodes and deep omnidirectional contacts (inspired by Medtronic 3389 see, e.g., \cite{anderson2018optimized}) are modelled according to their actual shape and size. 

Using the anatomical information of BST's ICBM152 template and ZI's FE mesh generator \cite{galaz2023multi}, we constructed a detailed multi-compartment tetrahedral FE mesh in which all subcomponents of the anatomical segmentation, cortical and subcortical tissues, were explicitly treated as individual compartments. The BST-2-ZI pipeline facilitated the integration of anatomical data from BST \cite{tadel2011brainstorm} into ZI \cite{he2020zeffiro}, enabling the generation this FE mesh. 

The successful generation of a high-resolution LF matrix with 100.000 source positions for this configuration demonstrated the feasibility of localizing deep sources such as thalamic activity using a whole-head model. Considering the dense placement of the omnidirectional DBS contacts, the  high 0.6 mm base FE resolution for the  mesh and refinement towards the DBS probe were essential features to obtain a sufficiently high mesh density  in the vicinity of the probe. 

The results of our source localization tests demonstrate that the presence of the stereotactically placed probe enhances the sensitivity of the lead field in its immediate vicinity. Sources aligned approximately parallel to the electrode axis were more accurately and focally reconstructed compared to those oriented perpendicularly. This effect is consistent with prior findings in DBS modeling and stereo-EEG studies, where the spatial distribution and intensity of electric fields have been shown to depend on electrode orientation \cite{li2021impact}. 

An obvious future work topic would be to use the present approach to generate an accurate model for some local environment of the brain, for example, a thalamic parcellation, which could be done in a straightforward manner using the existing code platform. Such a study could also cover an application-specific FE mesh adaptation to focus the model into a well-chosen region of interest, such as ANT which is targeted in various DBS studies. The findings concerning stereotactic source localization sensitivity motivates further numerical analysis concerning, for example, directional DBS probes, which have a complex pattern of electrodesthat required advanced FEM modeling. 

\section{Conclusion}

We have presented a novel pipeline integrating Brainstorm and Zeffiro Interface for constructing whole-head, multi-compartment volume conductor models suitable for stereo-EEG simulations. Our results demonstrate that combining non-invasive scalp electrodes with invasive stereo-EEG probe contacts using a multi-compartment finite element framework with complete electrode model boundary conditions can be advantageous in  source localization, particularly stereotactic applications targeting specific regions of the brain, such as the thalamus and other deep structures.

\bibliographystyle{IEEEtran}
\bibliography{sample.bib}

@book{knosche2022eeg,
  title={EEG/MEG Source Reconstruction},
  author={Kn{\"o}sche, Thomas R and Haueisen, Jens},
  year={2022},
  publisher={Springer}
}

@article{anso2022concurrent,
  title={Concurrent stimulation and sensing in bi-directional brain interfaces: a multi-site translational experience},
  author={Ans{\'o}, Juan and Benjaber, Moaad and Parks, Brandon and Parker, Samuel and Oehrn, Carina Renate and Petrucci, Matthew and Gilron, Ro’ee and Little, Simon and Wilt, Robert and Bronte-Stewart, Helen and others},
  journal={Journal of Neural Engineering},
  volume={19},
  number={2},
  pages={026025},
  year={2022},
  publisher={IOP Publishing}
}

@misc{brainstormFEM,
  author       = {{Brainstorm}},
  title        = {{FEM Mesh Generation}},
  howpublished = {\url{https://neuroimage.usc.edu/brainstorm/Tutorials/FemMesh}},
  note         = {Accessed: 2025-05-23},
  year         = {2025}
}

@misc{zeffiroBST2ZI,
  author       = {{Zeffiro Interface}},
  title        = {{BST\_2\_ZI – Brainstorm to Zeffiro Interface Conversion Guide}},
  howpublished = {\url{https://github.com/sampsapursiainen/zeffiro\_interface/wiki/BST\_2\_ZI}},
  note         = {Accessed: 2025-05-23},
  year         = {2025}
}

@article{MEDANI_2023_119851,
title = {Brainstorm-DUNEuro: An integrated and user-friendly Finite Element Method for modeling electromagnetic brain activity},
journal = {NeuroImage},
volume = {267},
pages = {119851},
year = {2023},
issn = {1053-8119},
doi = {https://doi.org/10.1016/j.neuroimage.2022.119851},
url = {https://www.sciencedirect.com/science/article/pii/S1053811922009727},
author = {Takfarinas Medani and Juan Garcia-Prieto and Francois Tadel and Marios Antonakakis and Tim Erdbrügger and Malte Höltershinken and Wayne Mead and Sophie Schrader and Anand Joshi and Christian Engwer and Carsten H. Wolters and John C. Mosher and Richard M. Leahy},
}

@article{hamalainen1993magnetoencephalography,
  title={Magnetoencephalography—theory, instrumentation, and applications to noninvasive studies of the working human brain},
  author={H{\"a}m{\"a}l{\"a}inen, Matti and Hari, Riitta and Ilmoniemi, Risto J and Knuutila, Jukka and Lounasmaa, Olli V},
  journal={Reviews of modern Physics},
  volume={65},
  number={2},
  pages={413},
  year={1993},
  publisher={APS}
}

@phdthesis{lahtinen2025mathematical,
  author       = {Joonas Lahtinen},
  title        = {Mathematical Methods for the Neuroelectromagnetic Inverse Problem for Focal and Unbiased Brain Imaging},
  school       = {Tampere University},
  year         = {2025},
  address      = {Tampere, Finland},
  type         = {Doctoral dissertation},
  isbn         = {978-952-03-3792-6},
  url          = {https://trepo.tuni.fi/handle/10024/187945}
}

@article{li2021impact,
  author    = {Li, Ang and Trotta, Nicholas and Worrell, Gregory A. and Yang, Lizhu and He, Bin},
  title     = {Impact of electrode orientation and location on stereo-EEG source imaging},
  journal   = {NeuroImage},
  volume    = {232},
  pages     = {117866},
  year      = {2021},
  doi       = {10.1016/j.neuroimage.2021.117866},
  url       = {https://www.sciencedirect.com/science/article/pii/S1053811921000239},
  publisher = {Elsevier}
}

@article{anderson2018optimized,
  title={Optimized programming algorithm for cylindrical and directional deep brain stimulation electrodes},
  author={Anderson, Daria Nesterovich and Osting, Braxton and Vorwerk, Johannes and Dorval, Alan D and Butson, Christopher R},
  journal={Journal of neural engineering},
  volume={15},
  number={2},
  pages={026005},
  year={2018},
  publisher={IOP Publishing}
}

@article{tadel2011brainstorm,
  title={Brainstorm: A user-friendly application for MEG/EEG analysis},
  author={Tadel, Fran{\c{c}}ois and Baillet, Sylvain and Mosher, John C and Pantazis, Dimitrios and Leahy, Richard M},
  journal={Computational intelligence and neuroscience},
  volume={2011},
  number={1},
  pages={879716},
  year={2011},
  publisher={Wiley Online Library}
}

@article{galaz2023multi,
  title={Multi-compartment head modeling in EEG: Unstructured boundary-fitted tetra meshing with subcortical structures},
  author={Galaz Prieto, Fernando and Lahtinen, Joonas and Samavaki, Maryam and Pursiainen, Sampsa},
  journal={Plos one},
  volume={18},
  number={9},
  pages={e0290715},
  year={2023},
  publisher={Public Library of Science San Francisco, CA USA}
}

@inproceedings{gaser2016cat,
  title     = {CAT--A Computational Anatomy Toolbox for the Analysis of Structural MRI Data},
  author    = {Gaser, Christian and Dahnke, Robert},
  booktitle = {Human Brain Mapping Conference},
  year      = {2016},
  address   = {Geneva, Switzerland},
  note      = {Available at: \url{http://www.neuro.uni-jena.de/cat/}}
}

@article{riviere2003brainvisa,
  title     = {BrainVISA: an extensible software environment for sharing multimodal neuroimaging data and processing tools},
  author    = {Rivière, Denis and Cointepas, Yann and Papadopoulos-Orfanos, Dimitri and Cachia, Arnaud and Poupon, Cyril and Mangin, Jean-François},
  journal   = {NeuroImage},
  volume    = {19},
  number    = {3},
  pages     = {1873--1884},
  year      = {2003},
  publisher = {Elsevier},
  doi       = {10.1016/S1053-8119(03)00112-5}
}

@incollection{ad2006civet,
  title     = {CIVET: Automated Pipeline for Measuring Cortical Thickness},
  author    = {Ad-Dab'bagh, Yasser and Lyttelton, Oliver and Muehlboeck, Jan-Sebastian and Lepage, Catherine and Einarson, Dan and Mok, Katherine and Ivanov, Olga and Vincent, R. Daniel and Lerch, Jason and Fombonne, Eric and Evans, Alan C.},
  booktitle = {Proceedings of the 12th Annual Meeting of the Organization for Human Brain Mapping},
  year      = {2006},
  address   = {Florence, Italy},
  note      = {Available at: \url{http://www.bic.mni.mcgill.ca/ServicesSoftware/CIVET}}
}

@article{shattuck2002brainsuite,
  title     = {Brainsuite: An automated cortical surface identification tool},
  author    = {Shattuck, David W. and Leahy, Richard M.},
  journal   = {Medical Image Analysis},
  volume    = {6},
  number    = {2},
  pages     = {129--142},
  year      = {2002},
  publisher = {Elsevier},
  doi       = {10.1016/S1361-8415(02)00054-3}
}

@article{ashburner2005unified,
  title     = {Unified segmentation},
  author    = {Ashburner, John and Friston, Karl J.},
  journal   = {NeuroImage},
  volume    = {26},
  number    = {3},
  pages     = {839--851},
  year      = {2005},
  publisher = {Elsevier},
  doi       = {10.1016/j.neuroimage.2005.02.018}
}

@article{dale1999cortical,
  title     = {Cortical surface-based analysis: I. Segmentation and surface reconstruction},
  author    = {Dale, Anders M. and Fischl, Bruce and Sereno, Martin I.},
  journal   = {NeuroImage},
  volume    = {9},
  number    = {2},
  pages     = {179--194},
  year      = {1999},
  publisher = {Academic Press},
  doi       = {10.1006/nimg.1998.0395}
}

@article{henschel2020fastsurfer,
  title     = {FastSurfer - A Fast and Accurate Deep Learning Based Neuroimaging Pipeline},
  author    = {Henschel, Lukas and Conjeti, Sailesh and Estrada, Sergio and Diers, Kai and Fischl, Bruce and Reuter, Martin},
  journal   = {NeuroImage},
  volume    = {219},
  pages     = {117012},
  year      = {2020},
  publisher = {Elsevier},
  doi       = {10.1016/j.neuroimage.2020.117012}
}

@article{lancaster2007bias,
  title={Bias between MNI and Talairach coordinates analyzed using the ICBM-152 brain template},
  author={Lancaster, Jack L and Tordesillas-Guti{\'e}rrez, Diana and Martinez, Michael and Salinas, Felipe and Evans, Alan and Zilles, Karl and Mazziotta, John C and Fox, Peter T},
  journal={Human brain mapping},
  volume={28},
  number={11},
  pages={1194--1205},
  year={2007},
  publisher={Wiley Online Library}
}

@article{vorwerk2014guideline,
  title={A guideline for head volume conductor modeling in EEG and MEG},
  author={Vorwerk, Johannes and Cho, Jae-Hyun and Rampp, Stefan and Hamer, Hajo and Kn{\"o}sche, Thomas R and Wolters, Carsten H},
  journal={NeuroImage},
  volume={100},
  pages={590--607},
  year={2014},
  publisher={Elsevier}
}

@article{salayev2006spike,
  title={Spike orientation may predict epileptogenic side across cerebral sulci containing the estimated equivalent dipole},
  author={Salayev, Kamran Ali and Nakasato, Nobukazu and Ishitobi, Mamiko and Shamoto, Hiroshi and Kanno, Akitake and Iinuma, Kazuie},
  journal={Clinical neurophysiology},
  volume={117},
  number={8},
  pages={1836--1843},
  year={2006},
  publisher={Elsevier}
}

@article{gramfort2011forward,
  title={Forward field computation with OpenMEEG},
  author={Gramfort, Alexandre and Papadopoulo, Th{\'e}odore and Olivi, Emmanuel and Clerc, Maureen},
  journal={Computational intelligence and neuroscience},
  volume={2011},
  number={1},
  pages={923703},
  year={2011},
  publisher={Wiley Online Library}
}

@article{mosher1999eeg,
  title={EEG and MEG: forward solutions for inverse methods},
  author={Mosher, John C and Leahy, Richard M and Lewis, Paul S},
  journal={IEEE Transactions on biomedical engineering},
  volume={46},
  number={3},
  pages={245--259},
  year={1999},
  publisher={IEEE}
}

@article{he2020zeffiro,
  title={Zeffiro user interface for electromagnetic brain imaging: a GPU accelerated FEM tool for forward and inverse computations in Matlab},
  author={He, Qin and Rezaei, Atena and Pursiainen, Sampsa},
  journal={Neuroinformatics},
  volume={18},
  pages={237--250},
  year={2020},
  publisher={Springer}
}

@article{rezaei2021reconstructing,
  title={Reconstructing subcortical and cortical somatosensory activity via the RAMUS inverse source analysis technique using median nerve SEP data},
  author={Rezaei, Atena and Lahtinen, Joonas and Neugebauer, Frank and Antonakakis, Marios and Piastra, Maria Carla and Koulouri, Alexandra and Wolters, Carsten H and Pursiainen, Sampsa},
  journal={Neuroimage},
  volume={245},
  pages={118726},
  year={2021},
  publisher={Elsevier}
}

@article{lahtinen2024standardized,
  title={Standardized hierarchical adaptive Lp regression for noise robust focal epilepsy source reconstructions},
  author={Lahtinen, Joonas and Koulouri, Alexandra and Rampp, Stefan and Wellmer, J{\"o}rg and Wolters, Carsten and Pursiainen, Sampsa},
  journal={Clinical Neurophysiology},
  volume={159},
  pages={24--40},
  year={2024},
  publisher={Elsevier}
}

@article{miinalainen2019realistic,
  title={A realistic, accurate and fast source modeling approach for the EEG forward problem},
  author={Miinalainen, Tuuli and Rezaei, Atena and Us, Defne and N{\"u}{\ss}ing, Andreas and Engwer, Christian and Wolters, Carsten H and Pursiainen, Sampsa},
  journal={NeuroImage},
  volume={184},
  pages={56--67},
  year={2019},
  publisher={Elsevier}
}

@article{pursiainen2012complete,
  title={Complete electrode model in EEG: relationship and differences to the point electrode model},
  author={Pursiainen, S and Lucka, F},
  journal={Physics in Medicine \& Biology},
  volume={57},
  number={4},
  pages={999},
  year={2012},
  publisher={IOP Publishing}
}

@article{pursiainen2017advanced,
  title={Advanced boundary electrode modeling for tES and parallel tES/EEG},
  author={Pursiainen, Sampsa and Agsten, Britte and Wagner, Sven and Wolters, Carsten H},
  journal={IEEE Transactions on Neural Systems and Rehabilitation Engineering},
  volume={26},
  number={1},
  pages={37--44},
  year={2017},
  publisher={IEEE}
}

\end{document}